%% file: repeaters_arxiv.tex
\documentclass[]{aa}
\usepackage{graphicx}
\usepackage[english]{babel}
\usepackage{amsmath,amssymb}
\usepackage{xcolor}
\usepackage{xspace}
\usepackage{array,multirow}
\usepackage[varg]{txfonts}
\usepackage{array}
\newcolumntype{H}{>{\setbox0=\hbox\bgroup}c<{\egroup}@{}}

\usepackage[]{hyperref}
\hypersetup{colorlinks=true, linkcolor=blue, citecolor=blue, urlcolor=blue}

\graphicspath{ {figs/} } 

\newcommand{\pccm}{{\ensuremath\mathrm{\,pc\,cm^{-3}}}}
\newcommand{\radm}{{\ensuremath\mathrm{\,rad\,m^{-2}}}}

\def\frba{FRB~121102\xspace}
\def\frbb{FRB~180814.J0422+73\xspace}

\makeatletter
\renewcommand*\aa@pageof{, page \thepage{} of \pageref*{LastPage}}
\makeatother

\begin{document}

\title{Repeating fast radio bursts with WSRT/Apertif}
\input{inc-authors.tex}

\date{Received date / Accepted date}

\abstract
{Repeating fast radio bursts (FRBs) present excellent opportunities to identify FRB progenitors and host environments, as well as decipher the underlying emission mechanism. Detailed studies of repeating FRBs might also hold clues to the origin of FRBs as a population.}
{We aim to detect bursts from the first two repeating FRBs: \frba (R1) and \frbb (R2), and characterise their repeat statistics. We also want to significantly improve the sky localisation of R2 and identify its host galaxy.}
{We use the Westerbork Synthesis Radio Telescope to conduct extensive follow-up of these two repeating FRBs. The new phased-array feed system, Apertif, allows covering the entire sky position uncertainty of R2 with fine spatial resolution in a single pointing. The data were searched for bursts around the known dispersion measures of the two sources. We characterise the energy distribution and the clustering of detected R1 bursts.}
{We detected 30 bursts from R1. The non-Poissonian nature is clearly evident from the burst arrival times, consistent with earlier claims. Our measurements indicate a dispersion measure of $563.5(2)\pccm$, suggesting a significant increase in DM over the past few years.
Assuming a constant position angle across the burst, we place an upper limit of 8\% on the linear polarisation fraction for the brightest burst in our sample. We did not detect any bursts from R2.}
{A single power-law might not fit the R1 burst energy distribution across the full energy range or widely separated detections. Our observations provide improved constraints on the clustering of R1 bursts. Our stringent upper limits on the linear polarisation fraction imply a significant depolarisation, either intrinsic to the emission mechanism or caused by the intervening medium, at 1400\,MHz that is not observed at higher frequencies. The non-detection of any bursts from R2, despite nearly 300\,hrs of observations, implies either a highly clustered nature of the bursts, a steep spectral index, or a combination of both assuming the source is still active. Another possibility is that R2 has turned off completely, either permanently or for an extended period of time.}

\keywords{radio continuum: general -- stars: neutron -- pulsars: general}

\maketitle

\section{Introduction}
\label{sec:introduction}
Fast radio bursts (FRBs) are transient, highly luminous events characterised by their short timescales of typically only a few milliseconds and dispersion measures (DMs) which are generally much larger than those expected from the Galactic electron density. These properties suggest FRBs to have originated from compact, highly energetic extra-galactic sources \citep{Lorimer07,Thornton13}. Despite extensive follow-up, the majority of the discovered FRBs have been found to be one-off events \citep{Petroff2015}. However, to date, 20~FRBs have been reported to exhibit repeat bursts \citep{spitler16,R2CHIME19,CHIME19c,Patek19,2020arXiv200103595F}. For a recent review of FRBs, see \citet{phl19}.

The repeat bursts from some FRBs enable studies of several FRB properties which are otherwise very hard to do for one-off sources. For example, deep follow-up of the repeating FRBs makes it possible to determine their sky positions with extremely high precision, identify the host galaxies and even associated persistent radio, optical or high-energy sources, if present. The localisation precision also helps in following-up any transient, multi-wavelength emission associated with the bursts. The repetition of bursts from the same source constrains FRB theory as well. A distinct constraint is that a cataclysmic event cannot produce repeating FRBs, and the underlying emission process should be able to sustain and/or repeat itself over considerably long periods of at least several years. The repetition also helps in much detailed investigations of the individual bursts, e.g., using coherently dedispersed high time and frequency resolution and/or over a wide frequency span.

At the time of observations used in this work, two FRBs were known to repeat --- \frba and \frbb \citep[hereafter R1 and R2, respectively,][]{spitler16,R2CHIME19}. R1 is precisely localised \citep{Chatterjee17,Marcote17} to a low-mass, low-metallicity dwarf galaxy \citep{Tendulkar-2017}, which has helped in theoretically exploring the potential progenitors. Detailed radio follow-up has uncovered several intriguing features of R1. One particularly noteworthy feature is the complex time-frequency structures noted in several individual bursts, in the form of nearly 250\,MHz wide frequency bands (at 1400\,MHz) drifting towards lower frequencies \citep{hss+19}. These bands are not caused by interstellar scintillation, and rather likely to be either intrinsic to the emission process or caused by exotic propagation effects like plasma lensing. Similar frequency bands, albeit without the drifting in some cases, have also been observed from a number of Galactic neutron stars 
(the Crab pulsar PSR~B0531+21, \citealt{Hankins16}; the Galactic Center Magnetar PSR~J1745$-$2900, \citealt{Pearlman18}; Magnetar XTE~J1810$-$197, \citealt{Maan19b}). 
However, any links between these galactic neutron stars and FRBs are as yet unclear, and require further study.
In this work we therefore focus on repeating-FRB pulse-energy distribution and repetition statistics, and compare them to those of pulsar giant pulses and bursts from soft gamma ray repeaters (SGRs) and magnetars.

The arrival times of R1 bursts are not well-described by a homogeneous Poisson process \citep{scholz16,orp18}, implying a clustered nature of the bursts. The clustering of the bursts has important implications for accurately determining the repeat rate as well as optimal observing strategies. It might also contain clues about the emission mechanism. Furthermore, R1 bursts exhibit nearly 100\% linear polarisation at 4500\,MHz and an exceptionally large, rapidly varying rotation measure
\citep[RM; $1.33-1.46\times10^5\radm$;][]{msh+18}.
This indicates that the R1 bursts are emitted in, or propagate through, an extreme and varying magneto-ionic environment. Due to the potential inter-channel depolarisation caused by the exceptionally large RM, the polarisation characteristics at 1400\,MHz are not fully known. A polarimetric characterisation at this frequency needs observations with reasonably narrow ($\lesssim$100\,kHz) frequency channels.

R2 was discovered using the pre-commissioning data from the CHIME telescope which operates in a frequency range of 400 to 800\,MHz. Some of the bursts from R2 showed strong similarities with those from {R1} in terms of the complex time-frequency structures. However, the uncertainties in the sky position \citep[$\pm4'$ and $\pm10'$ in RA and Dec, respectively][]{R2CHIME19} have limited any more detailed comparisons between the two repeating FRBs as well as extensive studies of the R2 bursts themselves. A precise localisation using an interferometer will enable detailed polarimetric and high resolution studies of the R2 bursts as well as probes of the host galaxy and any associated persistent radio or high-energy source.

In this work, we aim to characterise the 1400\,MHz polarisation and clustering nature of the bursts from R1, particularly for the bursts at the brighter end of the energy distribution, and localise R2 as well as study many of the above-mentioned aspects of both the repeating FRBs. For this purpose, we have utilised primarily the commissioning data from the new Apertif system on the Westerbork Synthesis Radio Telescope (WSRT). Using the large data sets acquired on R1 and R2, we present here the emission statistics of these two FRBs, over the highest pulse energies, and the longest timescales so-far reported.

In the following sections, we provide more details on the time-domain observing modes used in this work (Sect.~\ref{sec:apertif}) as well as observations and data reduction methods (Sect.~\ref{sec:obs_and_data_reduction}). We present and discuss our results obtained for R1 and R2 in Sects.~\ref{sec:121102} and \ref{sec:r2}, respectively. The overall conclusions are summarised in Sect.~\ref{sec:conclusion}.

\section{The Apertif observing modes}
\label{sec:apertif}

Aperture Tile in Focus (Apertif) is the new phased-array feed system installed on WSRT. It increased the field-of-view (FoV) to ${\sim}8.7\,$ square degrees, turning WSRT into an efficient survey instrument (\citealt{ovc10,2019NatAs...3..188A}; van Capellen et al. in prep). The system operates in the frequency range of $1130-1760\,$MHz with a maximum bandwidth of $300\,$MHz. The frequency resolution depends on the observing mode, as explained in the next subsections. Each of the WSRT dishes beam-forms the 121 receiver elements into up to 40 partly-overlapping beams on the sky (hereafter compound beams), each with a diameter of roughly $35\arcmin$ at $1400\,\mathrm{MHz}$. The data are then sent to the central system, which either operates as a correlator for imaging (cf. Adams et al. in prep), or as a beamformer for time-domain modes \citep{leeu14}.

The time-domain observing mode system exploits the wide FoV to search for new FRBs as well as to localise any poorly localised repeating FRBs. This mode is enabled by a back-end that is capable of detecting such highly dispersed events in \mbox{(quasi-)real-time}. We commissioned this system on pulsars and repeating FRBs (van Leeuwen et al. in prep). The following two time-domain modes were used to obtain the data presented here.

\subsection{Baseband mode}
\label{sec:sub:sc1}
In baseband mode, the central beam of up to ten telescopes is combined to obtain a high-sensitivity beam in one direction, with a resolution of $32\arcsec \times 35\arcmin$. The pulsar backend then either performs real-time pulsar folding, or records the raw voltages with a time resolution of $1.28\mathrm{\,\mu s}$ and frequency resolution of $0.78125\,$MHz. This allows for coherent dedispersion, as well as choosing an optimal trade-off between time and frequency resolution. During commissioning, we were at first limited to a bandwidth of $200\,$MHz and a single polarisation. The total sensitivity of the single-polarisation system is a factor $\sqrt{2}$ lower than that of the full dual-polarisation system. In early 2019 the system was upgraded to dual-polarisation and in March 2019 the bandwidth was increased to $300\,$MHz.

\subsection{Survey modes}
\label{sec:sub:sc4}
In survey mode, the system beamforms all 40 dish beams either coherently or incoherently. In coherent mode, voltage streams are combined across eight dishes with the 
appropriate complex weights. The resulting tied-array beams (TABs) are narrow (${\sim} 35\arcsec$) in East-West direction, but retain the dish resolution of ${\sim} 35\arcmin$ in North-South direction. In incoherent mode, intensity data are summed across dishes for all 40 compound beams individually. The incoherent-array beams (IABs) retain the full dish field-of-view, but at a sensitivity loss of $\sqrt{N_\mathrm{dish}}$ compared to the TAB mode. For survey mode data, both polarisations are summed. The resulting Stokes I data are stored to disk in filterbank format with a time resolution of $81.92\,\mathrm{\mu s}$ and frequency resolution of $0.1953125\,$MHz \citep{2017arXiv170906104M}. The bandwidth of the survey mode data is the same as that of the baseband mode data: 200\,MHz until March 2019, and 300\,MHz since then.

\section{Observations and data reduction}
\label{sec:obs_and_data_reduction}

\subsection{Observations}
\label{sec:sub:obs}
\begin{figure}
    \centering
    \includegraphics[width=\columnwidth]{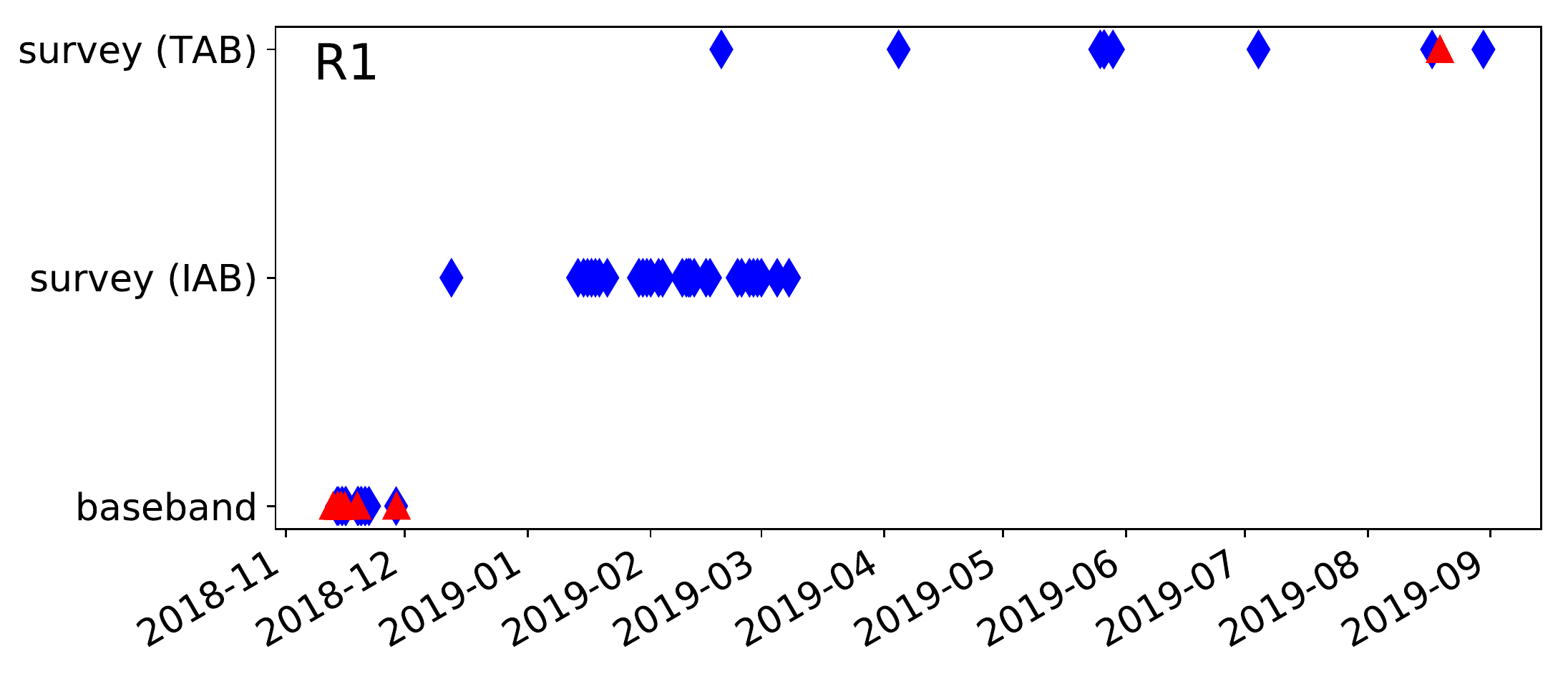} \\
    \includegraphics[width=\columnwidth]{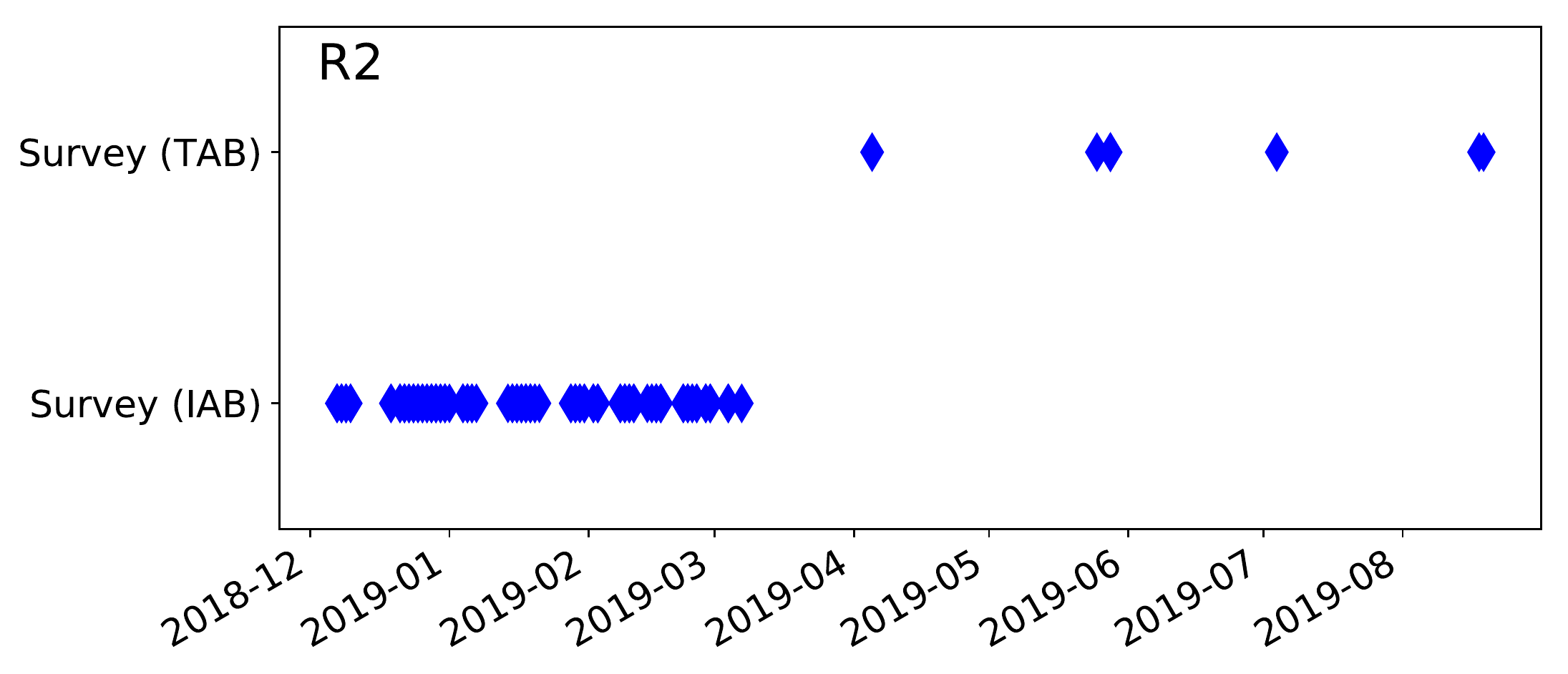} 
    \caption{Overview of Apertif observations of R1 (top) and R2 (bottom). Blue diamonds indicate observations without detected bursts, red triangles indicate observation with detected bursts. Along the vertical axis the observing mode is noted. R1 was observed for a total of ${\sim}130$ hours, and R2 for ${\sim}$300 hours.}
    \label{fig:obs}
\end{figure}

R1 and R2 were observed with Apertif between November 2018 and August 2019. R1 was observed in baseband mode, as well as both the incoherent and coherent survey modes. R2 is not well enough localised to be observed in the baseband mode, which is only suitable if the source is localised to within one Apertif tied-array beam. It was, however, observed with the incoherent and coherent survey modes. In total we spent ${\sim}130$ hrs on R1 and ${\sim}300$ hrs on R2. An overview of the observations is given in Fig.~\ref{fig:obs}.

\subsection{Reduction of R1 baseband data}
\label{sec:sub:analysis_sc1}
The baseband data were coherently dedispersed using the typical R1 DM of $560.5\pccm$ \citep{hss+19} and converted to filterbank using 
\texttt{digifil}. In this process, the time resolution was reduced from $1.28\,\mathrm{\mu s}$ to $51.2\,\mathrm{\mu s}$. This reduces 
the computation time while ensuring any burst of at least $51.2\,\mathrm{\mu s}$ in duration, less than the narrowest burst thus far reported, is still detectable. The filterbank data were then searched for any bursts with a DM between $520\pccm$ and $600\pccm$ in steps of $0.1\pccm$ with \texttt{PRESTO} \citep{presto}, using a threshold signal-to-noise (S/N) of 8. All candidates were visually inspected. The raw data of a 20 second window around each detected burst were saved for further analysis.

\subsection{Reduction of survey data}
\label{sec:sub:analysis_sc4}
For both survey modes, the data were analysed in real-time by our GPU pipeline, AMBER\footnote{\url{https://github.com/AA-ALERT/AMBER}} \citep{slb+16}. AMBER incoherently dedisperses the incoming Stokes I data to DMs between $0\pccm$ and $3000\pccm$ in steps of $0.2\pccm$ below $820\pccm$ and steps of $2.5\pccm$ above $820\pccm$, and writes a list of candidates with $\mathrm{S/N}\ge8$ to disk. These candidates are automatically further analysed by the offline processing mode of the ARTS processing pipeline, DARC\footnote{\url{https://github.com/loostrum/darc}}. First, the candidates are clustered in DM and time to identify bursts that were detected at multiple DMs or in multiple beams simultaneously. Of each cluster, only the candidate with the highest S/N is kept. For all remaining candidates, a short chunk of data (typically 2\,s) surrounding the candidate arrival time is extracted from the filterbank data on disk and dedispersed to the DM given by AMBER. These data are then given to a machine learning classifier \citep{cl18}, which determines whether a candidate is most likely a radio transient or local interference. For all candidates with a probability greater than $50\%$ of being an+ astrophysical transient, inspection plots are generated and e-mailed to the astronomers.

Because the pipeline was still in the commissioning phase, the data were also searched with a \texttt{PRESTO}-based pipeline as was done for the baseband data. For R1 we used the same DM range as for the baseband data. For R2, the DM range was set to $150 - 230
\pccm$, covering a wide range around the source DM of ${\sim}189\pccm$ \citep{R2CHIME19}. Both pipelines yielded identical results.

\section{R1}
\label{sec:121102}
\begin{figure*}
    \centering
    \includegraphics[width=\textwidth]{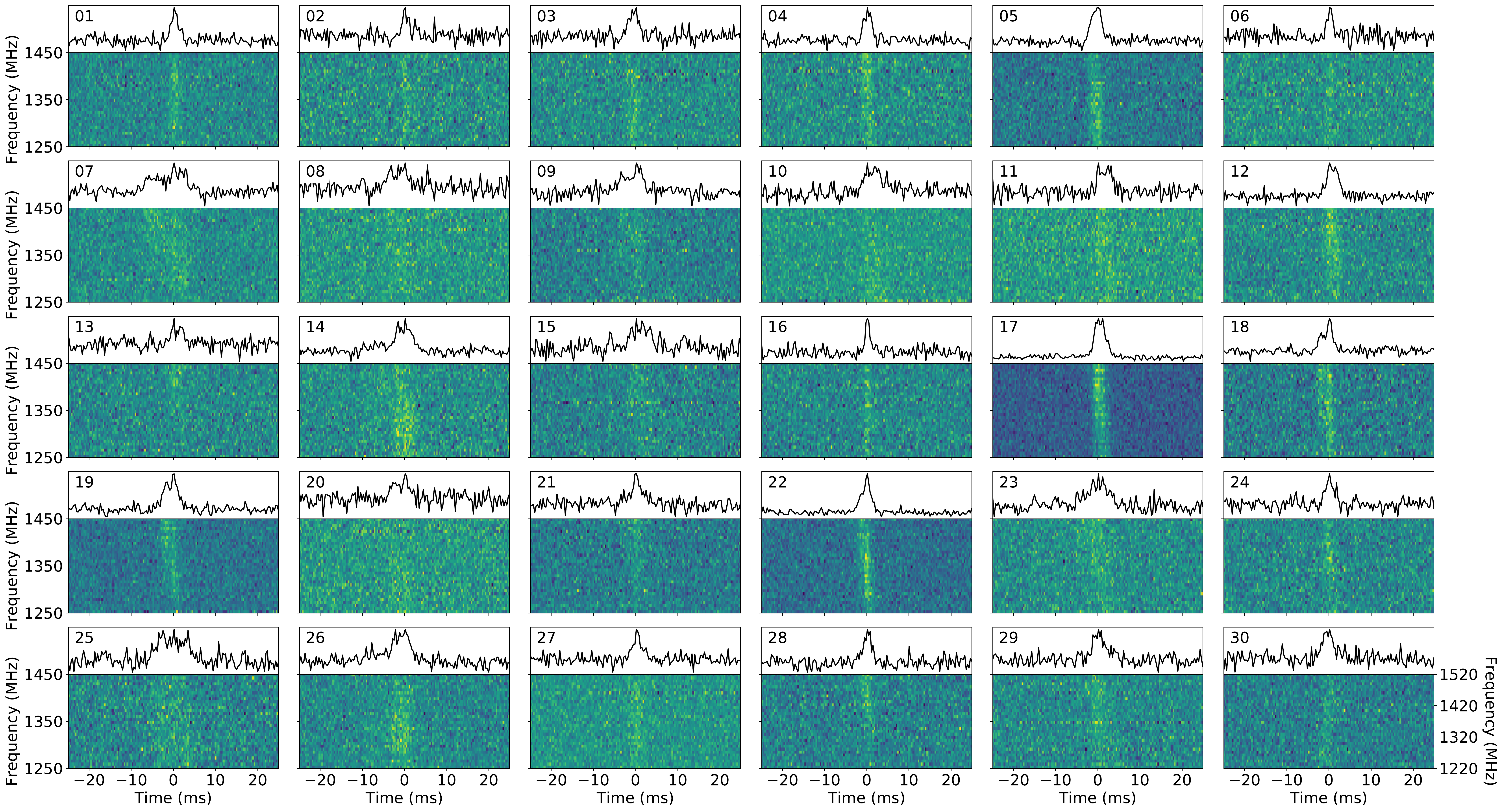}
    \caption{All 30 bursts from R1 detected with Apertif. Each burst was dedispersed to a DM of $560.5\pccm$. Bursts 1-29 were detected in offline searches of baseband data, while burst 30 was blindly found in real-time during TAB survey observations. All bursts are shown at a time resolution of $0.32768\,\mathrm{ms}$. Bursts 1-29 are shown at a frequency resolution of $6.25\,\mathrm{MHz}$ over a bandwidth of $200\,\mathrm{MHz}$, and were coherently dedispersed. Burst 30 was detected after a system upgrade to $300\,\mathrm{MHz}$ bandwidth and is shown at a frequency resolution of $9.375\,\mathrm{MHz}$, and is incoherently dedispersed. The residual intra-channel DM smearing in burst 30 is on the order of one sample and hence irrelevant. Larger versions of these plots are provided at \url{http://www.alert.eu/FRB121102/}.}
    \label{fig:burst_overview}
\end{figure*}

In total, 30 bursts were detected from R1. Of those, 29 were found in targeted baseband mode observations between 12 and 22 November 2018. One burst was found during regular TAB survey observations in August 2019. An overview of the bursts is shown in Fig.~\ref{fig:burst_overview}. 

\subsection{Flux calibration}
\label{sub:fluxcal}
The S/N of all bursts identified by the pipelines is determined using a matched filter with boxcar widths between 1 and 200 samples ($0.08$ to $16.4\,$ms). The noise is determined in an area around the burst visually confirmed to be free of interference. We used the modified radiometer equation \citep{CM03,MA14} to convert the obtained S/N to peak flux density. For an interferometer, the radiometer equation can be written as 
\begin{equation}
\label{eq:radiometer}
    S = \frac{\mathrm{S/N}\,T_\mathrm{sys}}{G N_\mathrm{dish}^{\,\beta} \sqrt{N_\mathrm{pol} \Delta\nu W}},
\end{equation}
where $S$ is the peak flux density, $T_\mathrm{sys}$ is the system temperature, $G$ is the gain of a single dish, $N_\mathrm{dish}$ is the number of dishes used, $\beta$ is the coherence factor, $N_\mathrm{pol}$ is the number of polarisations, $\Delta\nu$ is the bandwidth, and $W$ is the observed pulse width. We cannot readily measure $T_\mathrm{sys}$ and $G$ independently, but we can measure the system-equivalent flux density ($\mathrm{SEFD} = T_\mathrm{sys}/G$) of each dish. In order to do this, we performed drift scans of calibrator sources 3C147 and 3C286. The flux densities of both sources were taken from \citet{pb17}. In TAB mode, we typically find an SEFD of $700\,\mathrm{Jy}$ for the central beam of each dish. In addition, $\beta$ was shown to be consistent with $1$ for the TAB mode and with $1/2$ for the IAB mode \citep{2018PhDT.......155S}, as theoretically expected. These values were used to determine the sensitivity for each of the observations.
Although no bursts were detected in IAB mode, we can still use the radiometer equation to set an upper limit to the peak flux density of any bursts during those observations. 

For each detected burst, the peak flux density was converted to fluence by multiplying with the observed pulse widths, where the pulse width is defined as the width of a top-hat pulse with the same peak and integrated flux density as the observed pulse. This method of calculating the fluence is valid as long as the bursts are resolved in time, which all detected bursts are. Additionally we recorded the interval between that burst and the previous burst, or limits on the interval in case the burst was the first of an observation. 

An overview of the burst parameters is given in Table~\ref{tab:121102_overview}.

\begin{table*}
    \caption{Overview of parameters of the bursts detected from R1 with Apertif. The barycentric arrival times were calculated for a DM of $560.5\pccm$. Both the S/N-optimised and structure-optimised DM (where available) are shown. Details on how these were determined are given in Sect.~\ref{sub:dm}. The fluence was determined using the S/N-optimised DM. We assume a 20\% error on the fluences. The arrival times and wait times are typically accurate to a millisecond.}
    \label{tab:121102_overview}
    \centering
    \begin{tabular}{lllllll} \hline \hline
    Burst & Arrival time      & DM$_\mathrm{S/N}$ & DM$_\mathrm{struct}$ & Boxcar width & Fluence & Wait time\\
          & (barycentric MJD) & $\pccm$ & $\pccm$  & (ms)  & (Jy ms) &  (s) \\ \hline
	1 & 	58434.875313559 & 566(3) & & 2.3 & 3.9(8) & ${>}638.51$ \\
	2 & 	58434.889509584 & 566(3) & & 2.9 & 3.8(8) & 1226.537\\
	3 & 	58434.894775566 & 568(2) & & 2.9 & 4.1(8) & 454.980\\
	4 & 	58434.947599142 & 567(2) & & 2.6 & 5(1) & 4563.957\\
	5 & 	58434.966276186 & 567(2) & & 2.9 & 10(2) & 1613.701\\
	6 & 	58434.973174288 & 564(2) & & 2.3 & 3.1(6) & 595.991\\
	7 & 	58435.990826444 & 567(6) & & 6.9 & 27(5) & $1236.717$ -- $87925.148$\\
	8 & 	58436.040343161 & 570(4) & & 5.2 & 5(1) & 4278.245\\
	9 & 	58436.051467039 & 562(5) & & 6.2 & 10(2) & 961.100\\
	10 & 	58436.054576797 & 568(3) & & 4.3 & 6(1) & 268.683\\
	11 & 	58436.107672326 & 568(3) & & 3.6 & 6(1) & $3231.773$ -- $4587.543$\\
	12 & 	58436.110830017 & 568(2) & & 4.3 & 10(2) & 272.825\\
	13 & 	58436.121982835 & 565(2) & & 3.3 & 3.1(6) & 963.604\\
	14 & 	58436.123751681 & 566(4) & & 4.6 & 10(2) & 152.828\\
	15 & 	58436.132117748 & 568(4) & & 4.3 & 4.3(9) & 722.828\\
	16 & 	58436.192189901 & 564(1) & & 1.6 & 3.7(7) & $2433.671$ -- $5190.235$\\
	17 & 	58436.235117618 & 566(2) & 563.6$\pm$1.0 & 2.9 & 19(4) & 3708.959\\
	18 & 	58436.237531175 & 566(4) & 564.3$\pm$1.5 & 5.2 & 10(2) & 208.529\\
	19 & 	58436.242119138 & 565(4) & 563.9$\pm$1.8 & 3.9 & 10(2) & 396.399\\
	20 & 	58436.919260205 & 567(4) & & 4.6 & 5(1) & $3149.344$ -- $58504.989$\\
	21 & 	58436.964309370 & 569(2) & & 3.3 & 3.9(8) & 3892.246\\
	22 & 	58436.991334107 & 567(2) & 563.1$\pm$0.9 & 2.3 & 10(2) & $1276.105$ -- $2334.933$\\
	23 & 	58437.051643347 & 572(6) & & 3.9 & 8(2) & 5210.724\\
	24 & 	58437.899455547 & 564(2) & & 3.6 & 5(1) & $1433.866$ -- $73250.973$\\
	25 & 	58437.924610245 & 567(6) & & 8.2 & 9(2) & 2173.365\\
	26 & 	58437.993893047 & 561(4) & & 5.2 & 12(2) & $1492.858$ -- $5986.033$\\
	27 & 	58441.030569351 & 567(3) & & 4.3 & 5(1) & $4649.438$ -- $262368.833$\\
	28 & 	58450.903309478 & 566(2) & & 2.3 & 3.5(7) & $821.270$ -- $853004.746$\\
	29 & 	58450.974554110 & 565(2) & & 2.9 & 4.4(9) & 6155.538\\
	30 & 	58714.255429157 & 565(2) & & 2.6 & 5(1) & $1233.459$ -- $22747467.604$\\ \hline
    \end{tabular}
\end{table*}

\subsection{Dispersion measure}
\label{sub:dm}
The dispersion measure as determined by the pipelines (DM$_\mathrm{S/N}$) is optimised for S/N. The error on DM$_\mathrm{S/N}$ is calculated as the dispersion delay across the band that corresponds to half the pulse width. The bursts have complex frequency-time structure \citep{hss+19}. This is clearly visible in our sample as well, e.g. in bursts 18 and 19 (Fig.~\ref{fig:burst_overview}). While the S/N-optimised DM best captures the total energy output of the bursts, it is affected by the complex features that mimic dispersive effects. Though it is unclear whether these features are intrinsic or a propagation effect, their narrowband nature clearly distincts them from cold-plasma dispersion, which is described by a simple power law $\tau \propto \nu^{-2}$, where $\tau$ is the time delay at frequency $\nu$. The presence of these complex features led \citet{hss+19} to define a DM that maximises pulse structure (DM$_\mathrm{struct}$). As the subbursts drift downward in frequency, DM$_\mathrm{struct}$ is typically lower than DM$_\mathrm{S/N}$. As DM$_\mathrm{S/N}$ is based on the invalid assumption that the signal can be completely described by a $\nu^{-2}$ power law, while DM$_\mathrm{struct}$ is not, DM$_\mathrm{struct}$ is more likely to represent the actual dispersive effect.

Most of the Apertif bursts did not have a high enough S/N to reliably determine DM$_\mathrm{struct}$. We did attempt a fit for the brightest bursts with visually identifiable substructure. First, we set a DM by aligning the gaps between subbursts by eye. Then, we used {\sc dm\_phase}\footnote{\url{https://github.com/danielemichilli/DM_phase}} to fit the DM around this value. The error on the DM was taken as the maximum DM offset for which the coherent power diagnostic from {\sc dm\_phase} was at least half the maximum value.
For four bursts we were able to fit a DM$_\mathrm{struct}$ using this method. The values are listed in Table~\ref{tab:121102_overview}. We find an average DM$_\mathrm{struct}$ of $563.5(2)\pccm$.

This value is higher than the previously reported value of DM$_\mathrm{struct} = 560.57(7)\pccm$ \citep{hss+19}. We ran our pipeline on Apertif data of the Crab pulsar taken in November 2018 to verify the frequency labelling of our data. Several giant pulses were detected, all at the expected DM; we are thus confident the higher DM for this source is real. 
This increased DM of R1 as detected by Apertif is in line with an R1 burst detected by CHIME, with a DM of $563.6(5)\pccm$ \citep{CHIMER1}, and several bursts in Arecibo data (Seymour et al. in prep). These were all detected in the same week in November 2018 where most of the Apertif bursts were found.
In Fig.~\ref{fig:burst_overview}, where we display the bursts after dedispersion at the old value, this increase is already apparent in the residual dispersion slope. 

\begin{figure}
    \centering
    \includegraphics[width=\columnwidth]{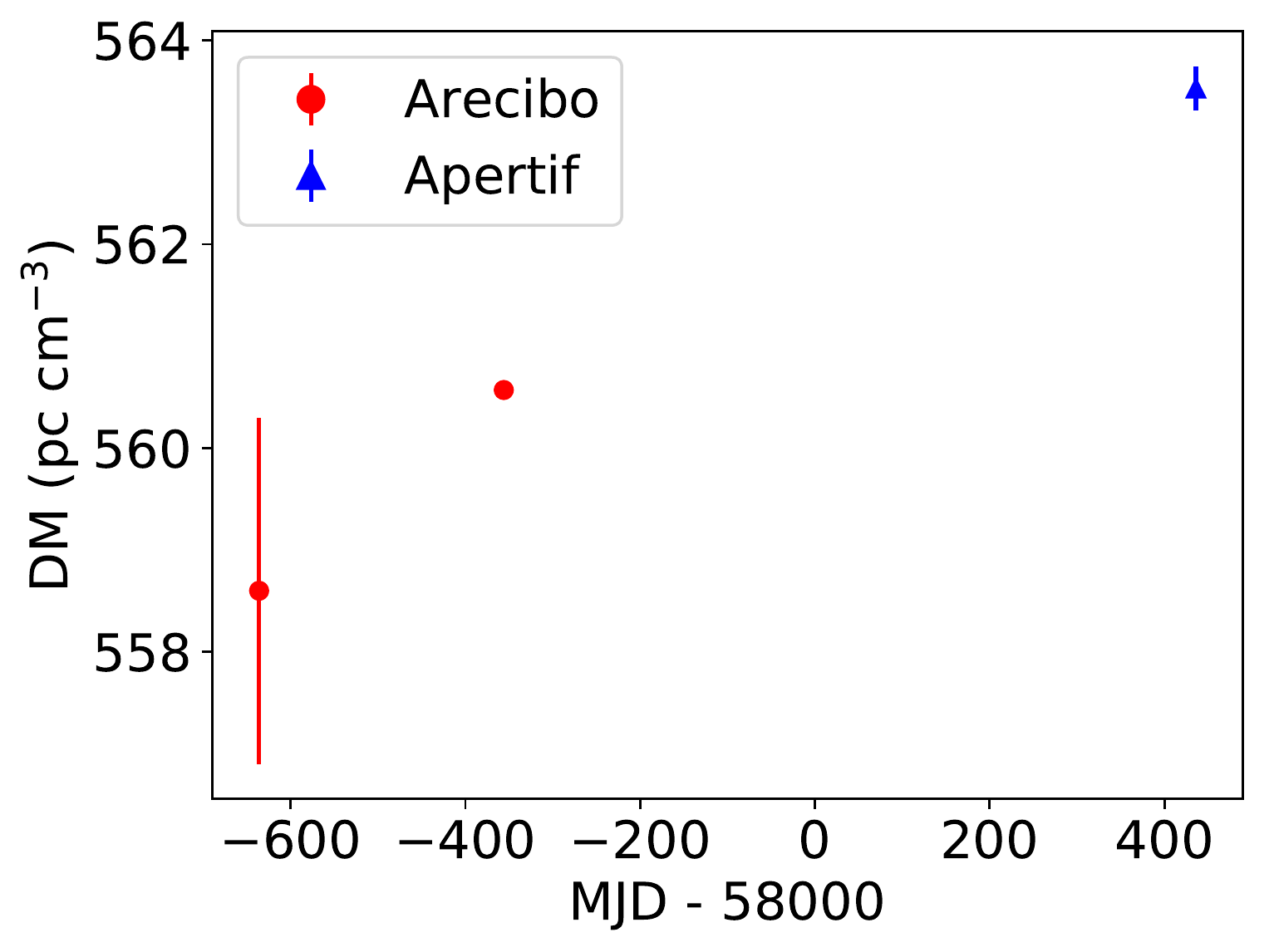}
    \caption{Structure-optimised DM of R1 at 1400\,MHz as measured at different epochs. The two Arecibo data points are from \citet{scholz16} and \citet{hss+19}, respectively. For Apertif, we averaged the DM of the four bursts for which we determined a structure-optimised DM. The error bars indicate 1$\sigma$ uncertainties. The uncertainty on the second data point is smaller than the circle. The DM has significantly increased from $560.57(7)\pccm$ in September 2016 to $563.5(2)$ in November 2018.}
    \label{fig:dm}
\end{figure}

In Fig.~\ref{fig:dm}, we show the observed DM$_\mathrm{struct}$ as observed at 1400\,MHz at different epochs. The significant increase in DM is clearly visible. As noted by \citet{CHIMER1}, the DM variation is likely to be local to the source, as such large changes are not seen in Galactic pulsars nor expected in the inter-galactic medium. It remains unclear whether these variations are stochastic or a secular trend \citep{hss+19,CHIMER1}. 
The increased DM, together with decreasing RM, show that R1 is in a highly magnetised, chaotic environment, where the RM and DM are unlikely to arise from the same region. The DM increase might be explained by a high-density filamentary structure moving into our line-of-sight, although several more complex models also predict changes in both DM and RM over time \citep[e.g.][]{piro18,metzger19}. 
In the supernova model of \citet{piro18}, the DM could increase during the Sedov-Taylor expansion phase. However, the rapid DM increase requires an age $\lesssim10^2\,$yrs while the DM is not expected to start rising until an age of $10^3-10^4\,$yrs. The decelerating blast wave model presented in \citet{metzger19} predicts stochastic DM changes, but typically at a lower level than observed here.
The rate at which the DM has increased thus remains hard to explain, although it may fit within the extremes of some models. The origin of this rapid change in DM will be an important aspect for future modelling efforts.

\subsection{Energy distribution}
\label{sub:energy_dist}
From the burst parameters, the intrinsic energy can be calculated as
\begin{equation}
    E = 4\pi\,d_L^2\,f_\mathrm{b}\,F\,\Delta\nu,
\end{equation}
were $E$ is the burst energy, $d_L$ is the luminosity distance \citep[$972\,$Mpc,][]{Tendulkar-2017}, $f_\mathrm{b}$ is the beaming fraction of the emission, $F$ is the fluence as observed on Earth, and $\Delta\nu$ is the intrinsic emission bandwidth. Following \citet{law17}, we assume isotropic emission, $f_\mathrm{b}=1$.

To investigate the energy distribution of R1 bursts, we consider the cumulative distribution of the mean burst rate, defined as the number of detected bursts divided by the total observing time \emph{including} observations without any detected bursts, as function of energy. It is known that the bursts show clustering in time (see \citealt{orp18} and Sect.~\ref{sub:rates}). Therefore it is important to consider the time scales probed by each set of observations: Ten observations spread over a year may yield very different results from ten identical observations spread over one week. Our data are supplemented with the data presented in
\citet{law17} and \citet{gourdji19}, who have performed similar analyses. An overview of the data used is shown in Table~\ref{tab:rates_data}. 

\begin{figure}
    \centering
    \includegraphics[width=\columnwidth]{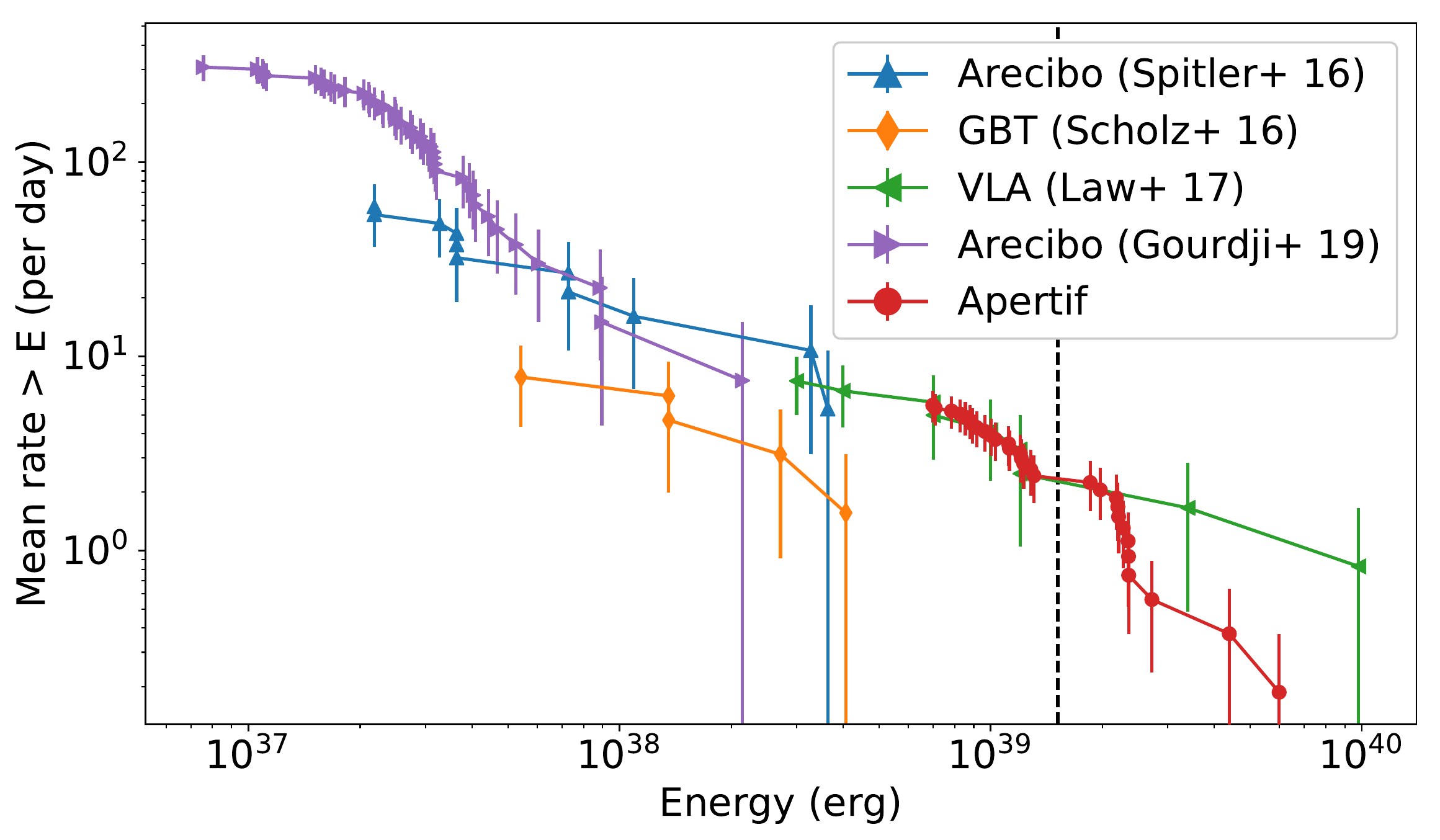}
    \caption{Cumulative distribution of R1 burst energies detected with VLA (3000\,MHz), GBT (2000\,MHz), Arecibo (1400\,MHz), and Apertif (1400\,MHz). The data used are described in Table~\ref{tab:rates_data}. 
    Poissonian errors on the rates are shown  for illustrative purposes. The typical power-law value for the early Arecibo, GBT, and VLA data is $-0.7$, while the later Arecibo data suggest a slope of $-1.8(3)$ above a completeness threshold of $2\times10^{37}\,$erg. The Apertif data suggest a slope of $-1.7(6)$ above the completeness threshold of $1.5\times10^{39}\,$erg, which is indicated by the vertical dashed line.}
    \label{fig:energy_dist}
\end{figure}

The resulting cumulative energy distributions are shown in Fig.~\ref{fig:energy_dist}. The distribution of burst energies has previously been characterised by a power law, $R({>}E) \propto E^{\gamma}$, where $R$ is the burst rate above energy $E$, and $\gamma$ is the power-law slope. \citet{law17} find a typical slope of $-0.7$ for VLA, GBT, and early Arecibo data. However, \citet{gourdji19} find a significantly steeper slope of $-1.8(3)$ using Arecibo data only, and suggest several reasons why the slope may be different. In some cases, the calculated burst energy is a lower limit. Moving some bursts to a higher energy would flatten the distribution. Additionally, the energies probed by the data presented in \citet{gourdji19} are lower than the others, where perhaps the slope is actually different or cannot be described by a power law at all. The slope is also strongly dependent on the chosen completeness threshold. Perhaps the different timescales probed by the different observations are also important. The 2016 Arecibo data used by \citet{gourdji19} (Table~\ref{tab:rates_data}) consist of two observations, one day apart, with detections of several bursts in each observation. This data set thus probes a relatively short timescale, which perhaps influences both the burst rate and energy distribution slope due to the clustered nature of the bursts.

To estimate the power-law slope of the Apertif burst energy distribution, we used Eq.~\ref{eq:radiometer} to set a completeness threshold for WSRT, using the threshold S/N of 8 and a typical pulse width of $4\,$ms. The least sensitive observations were using 8 dishes in IAB mode. The corresponding energy threshold is $1.5\times10^{39}\,$erg.
We then calculated the power-law slope using a maximum-likelihood estimator. If we include all data points, we find $\gamma=-1.3(3)$. When including only bursts above $1.5\times10^{39}\,$erg, the slope is $\gamma=-1.7(6)$. Although consistent with both $-0.7$ and $-1.8$ at the 2$\sigma$ level, the slope of the Apertif burst energy distribution favours the value found by \citet{gourdji19}. Although care must be taken in comparing slopes, this at least suggests that the probed energy range is not the reason for the steeper slope found in the 2016 Arecibo data that contain the lowest energy bursts, as with Apertif we are probing the highest burst energies at 1400\,MHz so far reported.

The most luminous Apertif burst presented here has an isotropic energy of ${\sim}4.5\times10^{39}\,$erg. During early commissioning, we reported the potential detection of a bright burst from R1 \citep{atel121102}. 
Its estimated isotropic energy was $1.2\times10^{40}\,$erg.
At that time, such bright bursts were not known to exist. 
The current energy distribution does, 
however,
credibly allow for a burst this bright.

Our slope $\gamma = -1.7(6)$ is the same as the power-law indices found by the studies listed in Table~\ref{tab:rates_data}, for Crab giant pulses
at the same observing frequency of 1400\,MHz.
Other studies, however, report steeper values (e.g., \citealt{2012ApJ...760...64M}; for an overview,
see \citealt{mikh18}).
The steepness is basically unchanged at frequencies a decade lower: at 150\,MHz it is still $-$2.04(3) \citep{lkk+19}.
The similarity between the brightness distribution fall-off seen in both FRBs and giant pulses suggests these could be related. 
In contrast, most regular pulsar emission follows a  log-normal intensity distribution  \citep[as discussed
  in e.g.][]{2002MNRAS.332..109J,2020MNRAS.tmp..131O}.

The energy distribution of the radio burst emission from magnetars could be significantly
different at different observing frequencies, and even at different spin phases. During its recent
outburst, the power-law indices for the radio bursts from magnetar XTE~J1810$-$197 span a
range between $-2.4(2)$ and $-0.95(30)$ \citep[0.65$-$1.36\,GHz;][]{Maan19b}. A similar range
was found in a study during its previous outburst \citep{Serylak09}. This observed range of power-law
indices for magnetar XTE~J1810$-$197 is consistent with our measurement of $\gamma$ for R1.

The distribution steepness matches less convincingly with a neutron-star emission mode that has also been put forward as a source model for FRBs: 
the soft gamma-ray bursts from magnetars \citep{2019ApJ...879....4W}, however if the FRB energy scales with voltage rather than Poynting flux, the distribution matches that of R1 more closely \citep{2019arXiv191006979W}.
The brightness distribution seen in SGR~1900+14 X-ray bursts is less steep, following a $-$0.66(13) trend \citep{1999ApJ...526L..93G}.

\begin{table*}
    \centering
    \caption{Data used to determine burst rates and energies for R1 at different epochs, and for reference the same data for the Crab pulsar, XTE\,J1810-197, and SGR~1900+14. The last column shows the derived power-law slope of the cumulative burst rate as function of energy.}
    \label{tab:rates_data}
    \renewcommand{\arraystretch}{1.4}
    \begin{tabular}{*6l}
    \hline 
    Source & Telescope & Bursts & T$_\mathrm{obs}$ (hr) & Date span  & Power-law index ($\gamma$) \\
    \hline
    \emph{R1} & Arecibo$^{(1,3)}$ & 11  &   4.5  & 2012-11-02 -- 2015-06-02 & $-$0.8$^{+0.3}_{-0.5}$    \\
      & GBT$^{(2,3)}$      & 5    &    15.3      & 2015-11-13 -- 2016-01-11 & $-$0.8$^{+0.4}_{-0.5}$    \\
      & VLA$^{(3)}$      & 9    &    28.9      & 2016-08-23 -- 2016-09-22 & $-$0.6$^{+0.2}_{-0.3}$    \\
      & Arecibo$^{(4)}$  & 41   &    3.2       & 2016-09-13 -- 2016-09-14 & $-$1.8(3) \\
      & WSRT$^{(5)}$     & 30   &    128.4     & 2018-11-12 -- 2019-08-30 & $-$1.7(6) \\ 
      \hline 
    \emph{Crab pulsar giant pulses} &
      WSRT$^{(6)}$     & 13,000   &     6 &        2005-12-10               &  $-$1.79(1)   \\
      &ATCA$^{(7)}$     & 700      &     3 &       2006-01-31               &  $-$1.33(14)  \\
      \hline 
    \emph{Magnetar XTE~J1810$-$197 } &
      GMRT$^{(8)}$ (650\,MHz)  & 5597    &     2.05 & 2018-12-18 -- 2019-02-17 &  $-$2.4(2)   \\      
     &GMRT$^{(8)}$ (1.36\,GHz) & 219     &     0.33 & 2019-02-17               &  $-$0.95(30)   \\      
      \hline 
    \emph{SGR 1900+14 X-ray bursts} &
      BATSE+RXTE$^{(9)}$     & 1,000 &     ${\sim}$50 &        1998-1999               &  $-$0.66(13)  \\
    \hline 
    \end{tabular}
    \tablebib{(1) \citet{spitler16}; (2) \citet{scholz16}; (3) \citet{law17}; (4) \citet{gourdji19};
    (5) This work; (6)~\citet{Karuppusamy-2010}; (7) \citet{Bhat-2008}; (8) \citet{Maan19b} (9)~\citet{1999ApJ...526L..93G}. 
    }
\end{table*}

\subsection{Burst repetition rate}
\label{sub:rates}
If repeating FRB burst rates follow Poissonian statistics, the distribution of wait times would be an exponential distribution. However, it has been shown that R1 bursts are highly clustered, which is incompatible with Poissonian statistics \citep{orp18}. A generalisation of the exponential distribution that allows for clustering is the Weibull distribution, defined as
\begin{equation}
\label{eq:wb}
\mathcal{W}(\delta|k,r) = \frac{k}{\delta} \, \left[\delta \, r\, \Gamma\left(1 + 1/k\right)\right]^k \, \mathrm{e}^{-\left[\delta \, r \, \Gamma\left(1 + 1/k\right)\right]^k},
\end{equation}
where $\delta$ is the burst interval, $r$ is the mean burst rate, $\Gamma(x)$ is the gamma function, and $k$ is a shape parameter. $k=1$ is equivalent to Poissonian statistics, $k<1$ indicates a preference for short burst intervals, i.e. bursts are clustered in time, and $k\gg1$ indicates a constant burst rate $r$. 

In order to apply the Weibull formalism to the R1 bursts observed with Apertif (cf. Table~\ref{tab:121102_overview}), we need to consider that subsequent observations may have correlated burst rates as some observations occurred in short succession. This requires small modifications to the equations presented by \citet{orp18}. We add a maximum burst interval to their equations to allow for correlated observations. A derivation of the modified equations is given in Appendix~\ref{app:weibull}.

\begin{figure}
    \centering
    \includegraphics[width=\columnwidth]{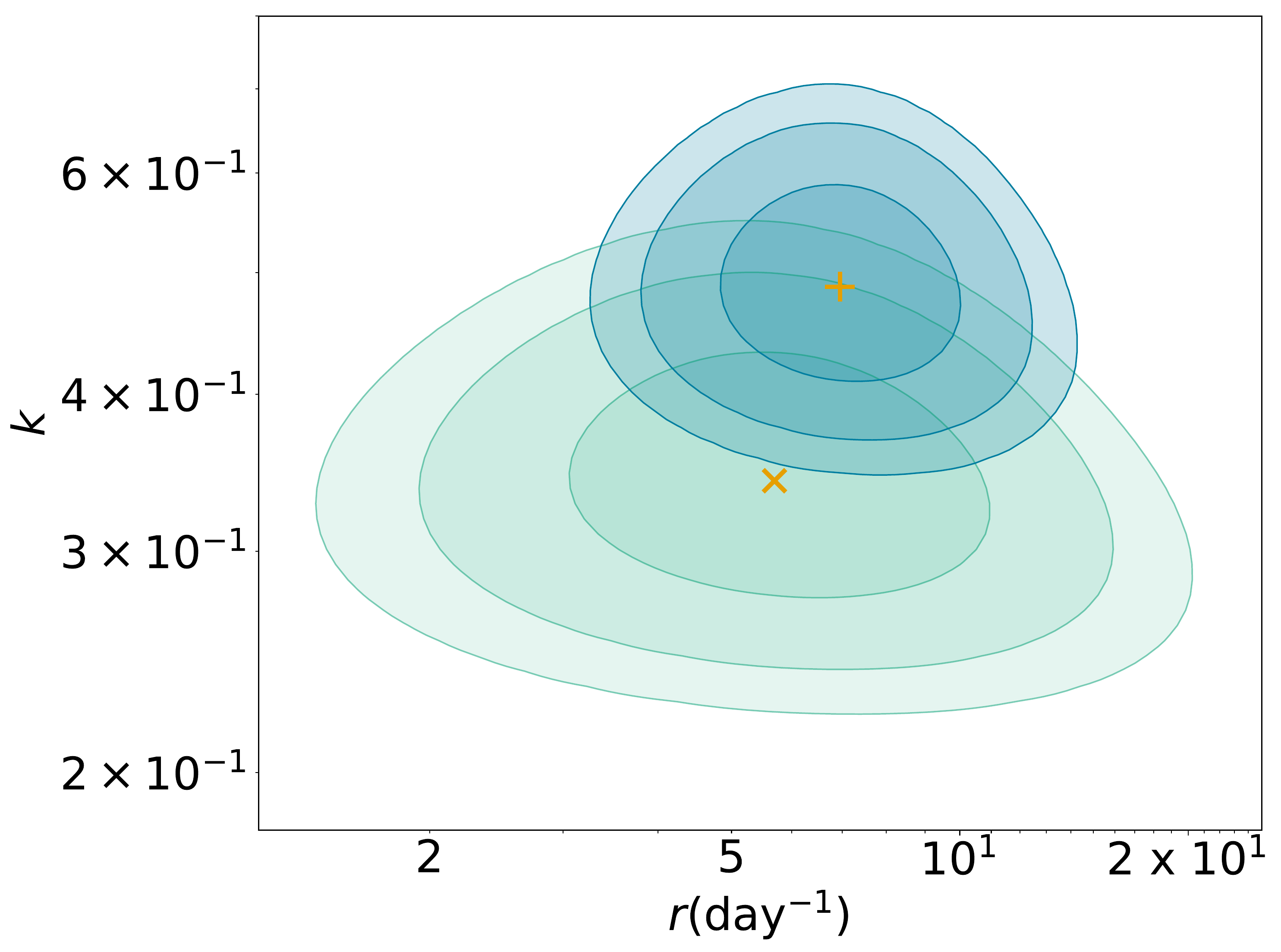}
    \caption{Posterior distribution of R1 burst rate and shape parameters. A \emph{lower} $k$ indicates a \emph{higher} degree of clustering. The green and blue areas indicate the result from \citet{orp18} and this work, respectively. The contours indicate 1, 2, and 3$\sigma$ limits on $r$ and $k$. The best-fit parameters are indicated by the cross and plus.}
    \label{fig:posterior_r1_wsrt}
\end{figure}

We assume a flat prior on both $k$ and $r$, only requiring that both are positive, and calculate the posterior as the product of the likelihoods of all Apertif observations. The posterior distribution is shown in Fig.~\ref{fig:posterior_r1_wsrt}. The best-fit parameters are $r=6.9^{+1.9}_{-1.5}\,\mathrm{day^{-1}}$ and $k=0.49^{+0.05}_{-0.05}$. Although \citet{orp18} used data from different instruments with different sensitivity thresholds, they all have a lower sensitivity threshold than Apertif. Given the negative slope of the energy distribution (Fig.~\ref{fig:energy_dist}), we had thus expected to find a lower rate with Apertif than the rate reported by \citet{orp18}. Given the uncertainties, the Apertif rate could still be lower, although we note that our best-fit rate and shape are consistent with that of \citet{orp18} at the $2\sigma$ level.

A Poissonian burst rate distribution ($k=1$) is excluded at high significance. This is not surprising, given that all bursts except one were detected within the first 30 observing hours out of a total of ${\sim}130$ hrs. The burst \emph{rate}, however, is consistent with the Poissonian estimate of $5.6(1)\,\mathrm{day^{-1}}$, even though Poissonian statistics cannot explain the distribution of burst \emph{intervals}. Thus, at the time scales probed by our data set, the clustering effect is not important in determining the average burst rate, but it does strongly influence the expected number of detected bursts for any single observation.

While the Weibull distribution does not fit previously observed R1 burst wait times very well, it is a significant improvement over Poissonian statistics \citep{orp18}. There are, however, other ways to describe the clustered behaviour R1 shows. For example, the burst rate might be described by several distinct Poisson processes: one (or more) with a high rate (the "active" state), and one (or more) with a low or zero rate (the "inactive" state). If the burst rate follows Poissonian statistics during an active period, i.e. there is only one burst rate during an active state, the wait time distribution is an exponential distribution. Samples of R1 wait times have indeed been shown to be consistent with exponential \citep{lin19}, but also with log-normal \citep{gourdji19} and power-law \citep{lin19} distributions during the active state, where in some cases it is not possible to distinguish between these distributions.

\begin{figure}
    \centering
    \includegraphics[width=\columnwidth]{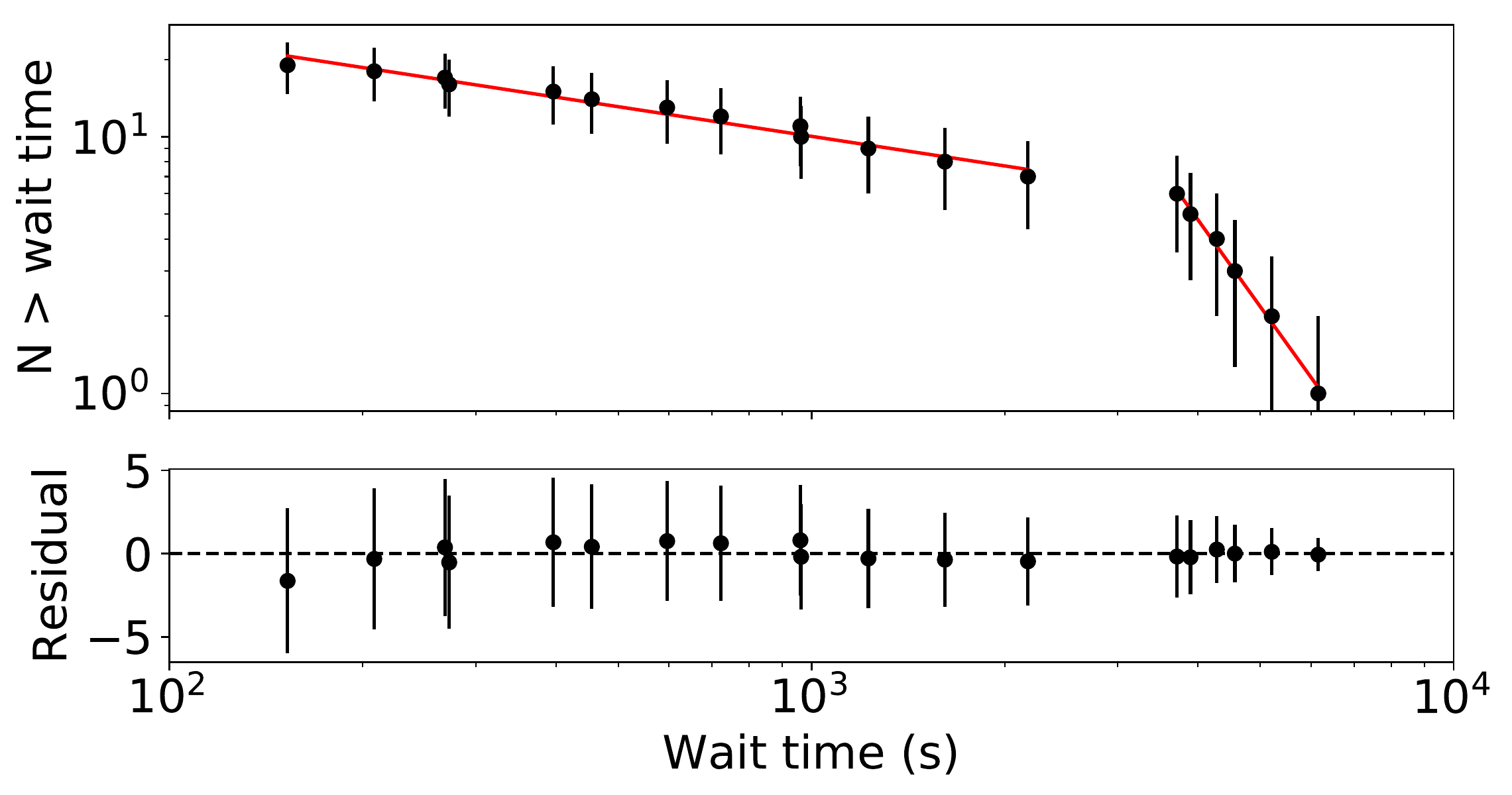}
    \caption{Wait time distribution of R1 bursts as detected by Apertif. Error bars are 1$\sigma$ Poissonian uncertainties.
    The distribution is bi-modal. The lower part can be fit by a power-law with slope $-0.38(2)$, but does not fit an exponential distribution. The higher part can be fit by either an exponential distribution or a power law with slope $-3.5(1)$. For both parts, the power-law fit is shown.}
    \label{fig:wait_times}
\end{figure}

Considering our observations in November 2018, where 29 out of the 30 bursts were detected, as the active state, we looked at the observed wait times during that time frame. In total, 19 wait times were determined (cf. Table~\ref{tab:121102_overview}). The resulting wait time distribution is shown in Fig.~\ref{fig:wait_times}. The Apertif sample shows a bi-modal wait time distribution. There is a dearth of burst times between ${\sim}2300\,$s and ${\sim}3600\,$s. The sample of wait times below $2300\,$s does not fit an exponential distribution, but can be fit with a power-law with slope $-0.38(2)$. The sample above $3600\,$s \emph{can} be fit by an exponential distribution, but the steep slope requires a Poissonian burst rate of ${>}25\,\mathrm{day^{-1}}$, which is incompatible with the observed rate.
However, it can be fit equally well with a power-law with slope $-3.5(1)$. In Fig.~\ref{fig:wait_times}, the best-fit power law is shown for both samples. Care has to be taken when interpreting these results, as there is a maximum wait time that can be detected in our observations of typically 2 hrs in duration. The chance of not detecting a given wait time increases linearly with the wait time, and any wait time larger than the observation duration is of course not observable. However, this effect cannot explain the bi-modality nor change in power-law index as those are non-linear in wait-time.

Our results indicate that during the active state in November 2018, the burst intervals did not follow a stationary Poisson process. This is incompatible with the wait time distribution of Crab giant pulses, which can be described by an exponential distribution \citep{L95}. However, a non-stationary Poisson process can result in a power-law wait time distribution at long wait times, which flattens towards shorter wait times. This is seen in for example X-ray solar flares \citep{aschwanden10,wheatland00}. The power-law slope depends on the exact form of the burst rate as function of time, but is generally flatter if the burst rate varies rapidly \citep{aschwanden10}.
In magnetar SGR~1900+14, the waiting time distribution between  bursts follows a log-normal function,
also indicative of a self-organized critical system  \citep{1999ApJ...526L..93G}.

\subsection{Polarisation properties}
\label{sub:pol}
Our baseband data came from only a single linear polarisation receptor, which makes it impossible to determine the total polarisation fraction. Even though the bursts from R1 are known to be highly linearly polarised, its high rotation measure \citep[RM > $10^5\radm$;][]{msh+18} implies that the polarisation angle sweeps around multiple times even within one Apertif frequency channel, so we do not expect to miss any bursts because of misalignment of the polarisation angle between a burst and the receiver elements. However, we can only estimate the RM and degree of linear polarisation \citep{ramkumar99,Maan15} if the depolarisation within a single channel is sufficiently small, which is clearly not the case at the native frequency resolution of Apertif.

Following \citet{msh+18}, the intra-channel polarisation angle rotation ($\Delta\theta$) is given by
\begin{equation}
    \Delta\theta = \frac{RM \mathrm{c}^2 \Delta\nu}{\nu^3},
\end{equation}
where $c$ is the speed of light, $\Delta \nu$ is the channel width and $\nu$ is the observing frequency. Evidently, $\Delta\theta$ is higher at lower frequencies, hence a much higher frequency resolution is required at 1400\,MHz than at 4500\,MHz. \citet{msh+18} find an intra-channel rotation of $~9\degr$, for a depolarisation fraction of $1.6\%$ for their data. At native frequency resolution, the Apertif data would be over $90\%$ depolarised. Therefore, we reprocessed the baseband data around the bursts and increased the number of channels to 4096 over a bandwidth of $200\,$MHz, implying a frequency resolution of ${\sim}49\,$kHz. This decreased the time resolution to $20.48\,\mathrm{\mu s}$, which is still sufficient to resolve the bursts. The resulting depolarisation fraction is $3\%$ for an RM of $10^5\radm$.

Following the procedure of \citet{Maan15}, we performed a discrete Fourier transform on the intensity spectra in the $\lambda^2$-domain, at each of the time samples in the bursts to obtain corresponding Faraday spectra. The Faraday spectrum represents linearly polarised power as a function of RM. 
We did not find any significant linearly polarised power at any of the trial RMs in the range $10^4 - 3.4{\times}10^5\radm$. However, due to the low S/N of individual samples within the bursts, we were sensitive to only a reasonably high degree of linear polarisation (50\% for the brightest burst, but $>$95\% for the other bursts). The R1 bursts are known to exhibit a constant polarisation position angle (PA) over the full burst duration \citep{msh+18}. To probe linearly polarised emission with higher sensitivity, we used the intensity spectra averaged over the entire burst widths which is valid if the PA is constant. We again did not detect any significant linearly polarised emission. At a 5$\sigma$ level detection threshold, our upper limits on the linearly polarised fractions for the 3 brightest bursts in our sample, burst numbers 17, 7 and 5 in Fig.~\ref{fig:burst_overview}, are 8\%, 14\% and 16\%, respectively. Our limits implicitly assume presence of a single Faraday screen between the source and the observer, which is supported by previous observations \citep{msh+18}. At 4500\,MHz, the linear polarisation fraction was measured to be close to 100\% \citep{msh+18}. Hence there must be some additional intrinsic or extrinsic depolarisation at 1400\,MHz to explain our non-detection.

\section{R2}
\label{sec:r2}
Despite several hundred hours of observations with equivalent or better 
sensitivity than reported in \citet{R2CHIME19},
no bursts from R2 were detected by Apertif. This is in contrast 
to the six bursts detected by CHIME in 23 hrs of R2 transits. 
Due to difficulty in measuring their time-dependent sensitivity, 
\citet{R2CHIME19} calculate R2's repetition rate with three 
bursts above their fluence completeness threshold of 13\,Jy\,ms, 
found in a total of 14 hrs of exposure. The least 
sensitive observations in our data set were performed with ten 
dishes in IAB mode, resulting in a fluence completeness 
threshold of $8.5 \sqrt{\frac{W}{10 \mathrm{\, ms}}}\,\mathrm{Jy\,ms}$, 
where $W$ is the pulse width. Most observations were more sensitive, 
meaning the limits we derive from our non-detection are conservative. 

Our ${\sim}$300 hrs of exposure corresponds to several years worth of CHIME transits, 
and yet Apertif detected no R2 bursts.
We offer two possible explanations, which are addressed independently.

\subsection{Temporal clustering}
R2 shows some striking similarities to R1. 
Beyond repetition, R2 also has distinct time/spectral structure, 
with a march-down in frequency of adjacent sub-pulses.
It may also exhibit non-Poissonian, or clustered, repetition. 
As has been previously noted, temporal clustering 
of bursts can drastically increase the probability of 
zero events being discovered in a given observation, 
even if the average repetition rate is high 
\citep{connor-2016b, orp18}. Therefore, it is possible 
that the reason Apertif did not detect R2 is that 
it is highly clustered. 

\begin{figure}
    \centering
    \includegraphics[width=\columnwidth]{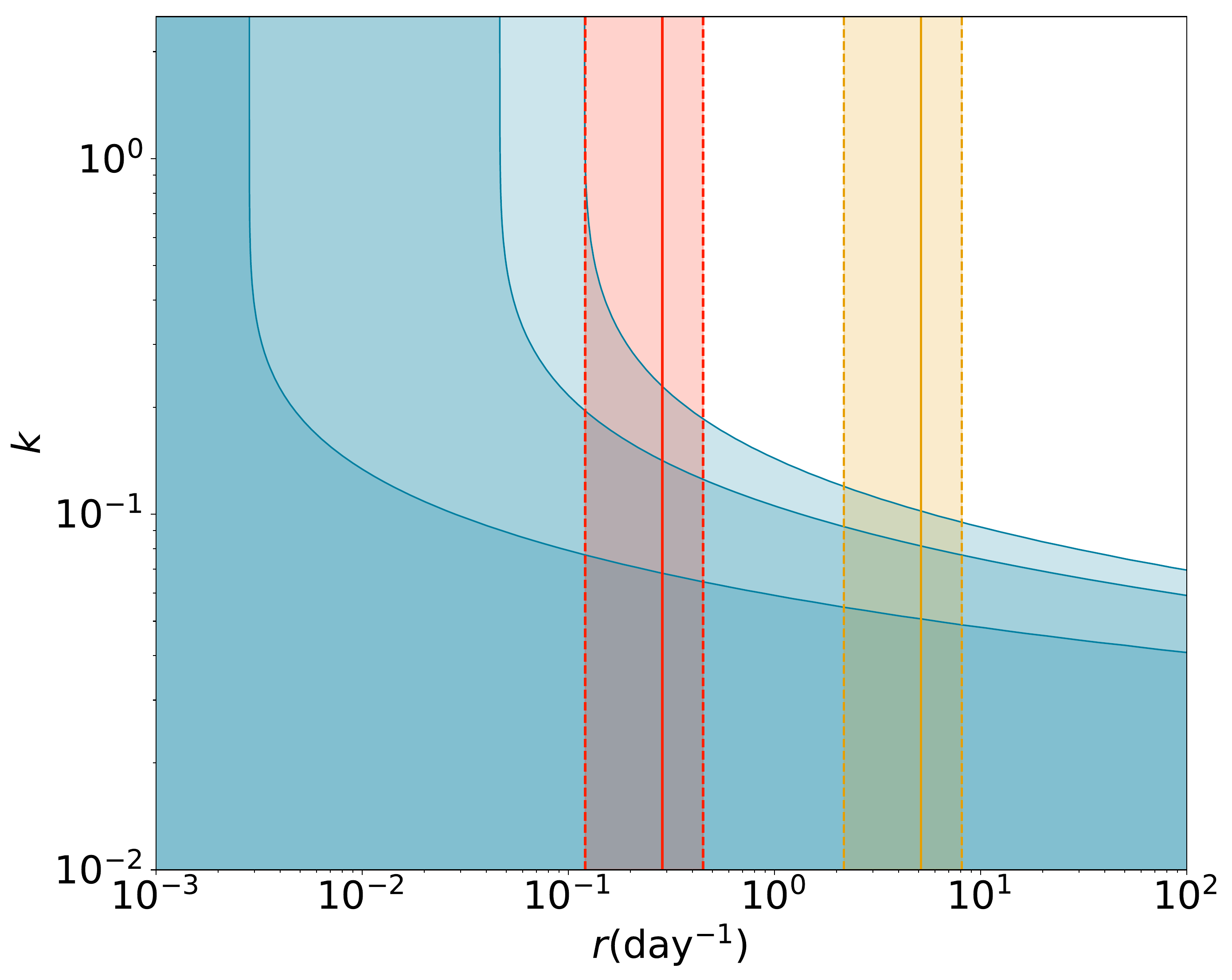}
    \caption{Posterior distribution of R2 burst rate and shape parameters. The contours indicate 1, 2, and 3$\sigma$ upper limits on $r$ and $k$. A \emph{lower} $k$ indicates a \emph{higher} degree of clustering. The yellow region indicates the CHIME rate with Poissonian error bars. The red region is the CHIME rate modified by a spectral index of -3.6, where the CHIME lower limit on the rate matches the Apertif upper limit at $k=1$, i.e. under the assumption of Poissonian statistics.}
    \label{fig:posterior_r2}
\end{figure}

The posterior for the burst rate $r$ and shape parameter $k$ 
of a Weibull distribution is shown in Fig.~\ref{fig:posterior_r2}, where small $k$ corresponds to high clustering. The burst rate as observed by CHIME \citep[>2.16 per day above 13 Jy ms][]{R2CHIME19} is indicated by the shaded yellow region. If we assume the same rate, $r$, of detectable 
pulses at CHIME and Apertif (i.e. a flat spectral index in repetition rate),
then within the Weibull framework, we can constrain the shape parameter, $k$, 
to be no greater than 0.12 at $3\sigma$. In other words, if R2's behaviour 
at 1400\,MHz and 600\,MHz are comparable, then the source's repetition 
has to be highly clustered, more so even than R1, for us 
not to detect any repeat bursts in ${\sim}300$ hrs of exposure. 

There are reasons to be sceptical of clustering as the sole explanation 
for our non-detection. For instance, if R2's repetition statistics 
were well-described by a Weibull distribution, then the
values of $k$ allowed by our non-detection 
imply that CHIME should have seen many bursts in a single transit, 
because the temporal clustering would be so significant. 
From a simple Monte Carlo simulation, 
we find that with $k\lesssim0.3$, half or more transits 
in which the FRB is seen to repeat should contain 
more than one repeat burst. Since CHIME saw its 
six repeat bursts in six distinct transits, $k$ is 
either not that small, or clustering only happens 
on longer time scales. Under our assumptions, 
the upper-bound on clustering set by our non-detection 
is inconsistent with the lower-bound on R2's clustering 
set by CHIME's observations. 

We also emphasise here that the Weibull distribution 
was chosen as a useful generalisation of the 
Poisson distribution, in order to account for 
the observed temporal clustering of R1. However, 
such clustering may not hold on all time scales, 
and FRB repetition wait times may not 
easily be described by a simple continuous distribution. Some FRBs 
may turn off entirely for extended periods, similar to 
X-ray binaries in quiescence, and then 
start back up with Poissonian repetition. Indeed, another 
explanation for our non-detection of R2 is that 
the source has turned off, either permanently or 
for a long, extended period. This will either 
be corroborated or falsified by CHIME's daily observing 
of R2 over the past year. 

\subsection{Frequency dependence}
If R2 is not significantly clustered, and 
the bursts follow Poissonian statistics ($k=1$), 
we set a $3\sigma$ upper limit to the burst 
rate at 1400\,MHz of $r<0.12$ 
per day above a fluence of $8.5\,$Jy ms. 
This limit is clearly inconsistent 
with the CHIME rate of $>$2.16 per day above 13\,Jy\,ms.
This indicates the source 
may be significantly less bright at 
1400\,MHz than at 600\,MHz. From the limited 
number of bursts detected 
by \citet{R2CHIME19}, it is difficult to assess the 
frequency-dependent rate of R2 and the authors do not 
provide a spectral index due to the banded nature of 
individual bursts. While the emission appears to 
occur at least over the full 400--800\,MHz CHIME band, 
three out of the five dynamic spectra shown for 
R2 appear to be confined within the bottom quarter of 
their frequency range \citep{R2CHIME19}. The source 
may therefore have a red spectrum.
Under the simplifying assumptions
that R2's pulses were always the same brightness at a given frequency, and were 
given by a power law across frequency such that $F(\nu)\propto \nu^{-\alpha}$,
then $\alpha$ must be greater than 3.6 at the 3$\sigma$ level
based on our data. 

However, it has become increasingly clear that bursts 
from repeating FRBs are given by bottom-heavy distributions, 
with many more dim events than bright ones \citep{gourdji19, CHIME19c}. The combination of 
frequency-dependence in the brightness of the source, as well as a power-law 
brightness distribution of repeat bursts (i.e. $F(\nu)\propto \nu^{-\alpha}$ \emph{and} $N(>F)\propto F^{-\gamma}$),
results in strong frequency-dependence in the detection rate. As we show in 
Appendix B, the frequency-dependent detection 
rate scales as $N(\nu) \propto \nu^{-\alpha\gamma}$, not as 
$N(\nu) \propto \nu^{-\alpha}$. In other words, 
if a source has a red spectrum ($\alpha>0$) and a steep brightness function,
then it will be difficult to detect at high frequencies. 
The analysis holds even if individual bursts from repeaters do 
not have power-law frequency spectra, so long as their average 
brightness as a function of frequency is a power law. 

This new effect may explain our non-detection of R2 at Apertif:
From the S/N listed in \citet{R2CHIME19}, $\gamma$\,$\approx$\,$2.2\pm1.3$, 
so even a moderate red frequency spectrum could result in considerably 
lower detection rates at 1400\,MHz vs 600\,MHz.

\section{Conclusions}
\label{sec:conclusion}
We have detected 30 bursts from R1 with Apertif. Their structure-optimised DM is higher than previously reported, consistent with an overall increase of ${\sim}2.7(2)\pccm\,\mathrm{yr^{-1}}$. The isotropic energy distribution of the bursts as determined by several instruments cannot be described by a single power law over the three decades of burst energies. The power-law slope as detected by Apertif, $\gamma = -1.7(6)$, is consistent with that of the Crab pulsar giant pulses and radio bursts from magnetar XTE\,J1810$-$197. Less convincingly it matches the X-ray bursts from magnetar SGR~1900+14. The repetition rate of the bursts matches with earlier found values, and confirms their highly clustered nature. Even when considering only the observations during an active period of the source, the burst arrival times are inconsistent with a stationary Poisson process and hence inconsistent with the wait time distribution of Crab giant pulses. However, the wait-time distribution can be described by a double power law, similar to solar flares. We place stringent upper limits on the linear polarisation fractions of some of the brightest bursts in our sample. For the brightest burst, the upper limit is $8\%$, assuming a constant polarisation angle across the burst. These limits suggest that there is an additional depolarising effect at 1400\,MHz that is not present at 4500\,MHz.

No bursts from R2 were detected. This might be because it has turned of either completely or for an extended period of time. If it has not turned off, the non-detection requires a high degree of clustering within the Weibull framework, assuming a flat spectral index. This is inconsistent with CHIME not having detected several bursts during one transit of R2. Alternatively, R2 may not emit in the Apertif band, or its emission may be intrinsically fainter. We find it unlikely that R2's statistical frequency spectrum can be described by a power-law. If it can, the spectral index has to be at least $\alpha>3.6$ to explain the Apertif non-detection.

\begin{acknowledgements}
\input{inc-acks.tex}
This work makes use of data from the Apertif system installed at the Westerbork Synthesis Radio Telescope owned by ASTRON. ASTRON, the Netherlands Institute for Radio Astronomy, is an institute of NWO.
\end{acknowledgements}

\bibliographystyle{yahapj}
\bibliography{journals_apj,psrrefs,modjoeri,modrefs,repeaters,yreferences}

\begin{appendix}
\section{Burst wait time formalism}
\label{app:weibull}
We describe the FRB wait time distribution by a Weibull distribution, following \citet{orp18}. The Weibull distribution is described by two parameters: the burst rate $r$ and clustering parameter $k$. $k=1$ is equivalent to Poissonian statistics, while a value much smaller or great indicate clustering in time and a constant burst rate, respectively.
We incorporate that subsequent observations can be correlated. Here we derive the modifications to the equations presented by \citet{orp18}.

The probability of measuring some set of burst arrival times $t_\mathrm{1}, t_\mathrm{2}, \dots t_\mathrm{N}$ in a single observation of duration $T$ can be split into three parts:
\begin{enumerate}
    \item The probability of the interval between the start of the observation and the first burst: $P(t_1)$
    \item The probabilities of the intervals between subsequent bursts in a single observation: $P(t_2 \dots t_\mathrm{N}) = \prod_\mathrm{i=1}^\mathrm{N-1} P(t_\mathrm{i+1} - t_\mathrm{i})$
    \item The probability of the interval between the last burst and the end of the observation: $P(T - t_\mathrm{N})$
\end{enumerate}

Assuming different observations are not correlated, points 1) and 3) describe minimum burst intervals. 
To include that subsequent observations can be correlated, we include a maximum burst interval, which is simply the interval between the last burst of an observation and the arrival time of next observed burst. Only for the intervals before the first detected burst and after the last burst, there is no constraint on the maximum burst interval. 

The addition of a maximum burst interval ($\delta_\mathrm{max}$) leads to several minor changes in the probability density functions of \citet{orp18}. 
The probability density distribution of the interval between the start of the observation and the first burst \citep[Eq.~13 of][]{orp18} is given by
\begin{equation}
\label{eq:wb_pdf_start}
\begin{aligned}
    &\mathcal{P}(t_1,\delta_\mathrm{max}|k,r) = r \, \int_{t_1}^{\delta_\mathrm{max}} \mathcal{W}(\delta|k,r) \, \mathrm{d}\delta \\
    &= r \, \left[\mathrm{CCDF}(t_1|k,r) - \mathrm{CCDF}(\delta_\mathrm{max}|k,r)\right],
\end{aligned}
\end{equation}

where $\delta$ is the interval between the last unobserved burst and the first observed burst, and CCDF is the cumulative complementary distribution function, defined as
\begin{equation}
\label{eq:wb_ccdf}
    \mathrm{CCDF}(\delta|k,r) = \int_\delta^\infty  \mathcal{W}(\delta^\prime|k,r) \, \mathrm{d}\delta^\prime = \mathrm{e}^{-\left[\delta \, r \, \Gamma(1+1/k)\right]^k}.
\end{equation}

The probability density of the intervals between subsequent bursts in a single observation is unchanged by our addition of correlated observations, and simply given by a product of Weibull distributions for the given intervals,
\begin{equation}
\label{wb_pdf_intervals}
\mathcal{P}(t_1|k,r) = \prod_\mathrm{i=1}^\mathrm{N-1} \mathcal{W}(t_{\mathrm{i}+1} - t_\mathrm{i}).
\end{equation}

The probability density of the interval between the last burst and the end of the observation is changed in a similar way as Eq.~\ref{eq:wb_pdf_start} and given by
\begin{equation}
\label{eq:wb_pdf_end}
\begin{aligned}
    &\mathcal{P}(T-t_\mathrm{N},\delta_\mathrm{max}|k,r) = \int_{T-t_\mathrm{N}}^{\delta_\mathrm{max}} \mathcal{W}(\delta|k,r) \, \mathrm{d}\delta\\
    &= \mathrm{CCDF}(T-t_\mathrm{N}|k,r) - \mathrm{CCDF}(\delta_\mathrm{max}|k,r).
\end{aligned}
\end{equation}

Lastly, we need to consider an observation without any detected bursts \citep[Eq.~17 of ][]{orp18}. The probability density distribution of such an observation is given by 
\begin{equation}
\label{eq:wb_pdf_0}
\begin{aligned}
    &P(N=0,\delta_\mathrm{max}|k,r) = r \int_T^{\delta_\mathrm{max}} \mathrm{CCDF}(t_1|k,r) \, \mathrm{d}t_1 \\
    &= \frac{\Gamma_\mathrm{i}{\left(1/k, \left(T\,r\,\Gamma(1+1/k)\right)^k\right)}}{k\,\Gamma{\left(1+1/k\right)}} \, - \, \\
    &\frac{\Gamma_\mathrm{i}{\left(1/k, \left(\delta_\mathrm{max}\,r\,\Gamma(1+1/k)\right)^k\right)}}{k\,\Gamma{\left(1+1/k\right)}},
\end{aligned}
\end{equation}
where $\Gamma_i(x, z)$ is the upper incomplete gamma function.
Note that in the limit $\delta_\mathrm{max} \rightarrow \infty$, all modified equations return to their equivalent versions for non-correlated observations.

\section{Frequency-dependent detection rate}

\noindent Suppose an FRB emits broad-band bursts with a power-law in frequency, given by
$L(\nu)\propto \left(\frac{\nu}{\nu_0}\right)^{-\alpha}$. 
If we assume the differential luminosity function of an 
individual repeater is given by a power-law $N(L)\propto L^{-(1+\gamma)}$, 
then the number of events above some minimum detectable luminosity is

\begin{equation}
    N(>\!L_{\mathrm{min}}) \propto \int\displaylimits_{L_{\mathrm{min}}}^{\infty} N(L)\,dL,
\end{equation}

\noindent where $L_\mathrm{min}$ is determined by the detection instrument's 
brightness threshold and the source's distance scale, such that $L_{\mathrm{min}}=4\pi d^2 S_{\mathrm{min}}$. 
If we then include the fact that the source is $\left(\frac{\nu}{\nu_0}\right)^{-\alpha}$ times brighter 
at frequency $\nu$ than at $\nu_0$, we find that $L_{\mathrm{min}}$ 
is decreased by that same factor, so

\begin{equation}
    N(>\!L_{\mathrm{min}}, \nu) \propto \int\displaylimits_{L_{\mathrm{min}}(\nu)}^{\infty} L^{-(1+\gamma)}\,dL.
\end{equation}

\noindent For $\gamma>0$, 

\begin{equation}
    N(>\!L_{\mathrm{min}}, \nu) \propto \left [ \left(\frac{\nu}{\nu_0}\right)^\alpha L_{\mathrm{min}} \right ]^{-\gamma},\, 
\end{equation}

\noindent and we find a strong relationship between observed repeat rate, $N(>\!L_{\mathrm{min}}, \nu)$, 
the source's spectral index $\alpha$, and its luminosity function index $\gamma$, such that

\begin{equation}
N(>\!L_{\mathrm{min}}, \nu) \propto \nu^{-\gamma\alpha}.
\end{equation}

\noindent This is striking, because it means that if a repeating FRB's brightness distribution 
deviates from $\gamma\approx1$, the source's detectability across frequency is 
significantly different from its \textit{brightness} across frequency. As an example, 
if R2 has $\gamma=2$, similar to the Crab, and $L(\nu)\propto \left(\frac{\nu}{\nu_0}\right)^{-2}$, there will be almost 30 
times fewer detectable bursts in the middle of the Apertif band vs. the middle of the CHIME band, 
assuming $S_{\mathrm{min}}$ is the same at both telescopes.

\end{appendix}

\end{document}

%% file: inc-authors.tex

\author{
     L.~C.~Oostrum        \inst{\ref{astron} \and \ref{uva}}
     \thanks{E-mail: l.c.oostrum@uva.nl}
\and Y.~Maan          \inst{\ref{astron}}
\and J.~van~Leeuwen       \inst{\ref{astron} \and \ref{uva}}
\and L.~Connor          \inst{\ref{uva} \and \ref{astron}}
\and E.~Petroff           \inst{\ref{uva} \and \ref{veni}}
\and J.~J.~Attema         \inst{\ref{escience}}
\and J.~E.~Bast           \inst{\ref{astron}}
\and D.~W.~Gardenier       \inst{\ref{astron} \and \ref{uva}}
\and J.~E.~Hargreaves     \inst{\ref{astron}}
\and E.~Kooistra          \inst{\ref{astron}}
\and D.~van~der~Schuur     \inst{\ref{astron}}
\and A.~Sclocco           \inst{\ref{escience}}
\and R.~Smits             \inst{\ref{astron}}
\and S.~M.~Straal        \inst{\ref{nyuad} \and \ref{nyuadcfa}}
\and S.~ter~Veen        \inst{\ref{astron}}
\and D.~Vohl            \inst{\ref{astron}}
\and E.~A.~K.~Adams       \inst{\ref{astron} \and \ref{kapteyn}}
\and B.~Adebahr           \inst{\ref{airub}}
\and W.~J.~G.~de~Blok     \inst{\ref{astron} \and \ref{cpt} \and \ref{kapteyn}}
\and R.~H.~van~den~Brink  \inst{\ref{astron} \and \ref{tricas}}
\and W.~A.~van~Cappellen   \inst{\ref{astron}}
\and A.~H.~W.~M.~Coolen   \inst{\ref{astron}}
\and S.~Damstra           \inst{\ref{astron}}
\and G.~N.~J.~van~Diepen  \inst{\ref{astron}}
\and B.~S.~Frank          \inst{\ref{sarao} \and \ref{cpt}}
\and K.~M.~Hess           \inst{\ref{astron} \and \ref{kapteyn}}
\and J.~M.~van~der~Hulst  \inst{\ref{kapteyn}}
\and B.~Hut               \inst{\ref{astron}}
\and M.~V.~Ivashina        \inst{\ref{chalmers}}
\and G.~M.~Loose          \inst{\ref{astron}}
\and D.~M.~Lucero         \inst{\ref{virginiatech}}
\and \'A.~Mika             \inst{\ref{astron}}
\and R.~H.~Morganti       \inst{\ref{astron} \and \ref{kapteyn}}
\and V.~A.~Moss           \inst{\ref{csiro} \and \ref{sydney} \and \ref{astron}}
\and H.~Mulder            \inst{\ref{astron}}
\and M.~J.~Norden         \inst{\ref{astron}}
\and T.~A.~Oosterloo      \inst{\ref{astron} \and \ref{kapteyn}}
\and E.~Orr\'u            \inst{\ref{astron}}
\and J.~P.~R.~de~Reijer   \inst{\ref{astron}}
\and M.~Ruiter            \inst{\ref{astron}}
\and N.~J.~Vermaas        \inst{\ref{astron}}
\and S.~J.~Wijnholds      \inst{\ref{astron}}
\and J.~Ziemke            \inst{\ref{astron} \and \ref{rugcit}}
} 

\institute{ASTRON, the Netherlands Institute for Radio Astronomy, Oude Hoogeveesedijk 4,7991 PD Dwingeloo, The Netherlands\label{astron}
  \and
Anton Pannekoek Institute, University of Amsterdam, Postbus 94249, 1090 GE Amsterdam, The Netherlands\label{uva}
  \and
Veni Fellow\label{veni}
  \and
Netherlands eScience Center, Science Park 140, 1098 XG, Amsterdam, The Netherlands\label{escience}
  \and
NYU Abu Dhabi, PO Box 129188, Abu Dhabi, United Arab Emirates\label{nyuad}
  \and
Center for Astro, Particle, and Planetary Physics (CAP$^3$), NYU Abu Dhabi, PO Box 129188, Abu Dhabi, United Arab Emirates\label{nyuadcfa}
  \and
Kapteyn Astronomical Institute, PO Box 800, 9700 AV Groningen, The Netherlands\label{kapteyn}
  \and
Astronomisches Institut der Ruhr-Universit\"at Bochum (AIRUB), Universit\"atsstrasse 150, 44780 Bochum, Germany\label{airub}
  \and
Dept.\ of Astronomy, Univ.\ of Cape Town, Private Bag X3, Rondebosch 7701, South Africa\label{cpt}
  \and
Tricas Industrial Design \& Engineering, Zwolle, The Netherlands\label{tricas}
  \and
South African Radio Astronomy Observatory (SARAO), 2 Fir Street, Observatory, 7925, South Africa\label{sarao}
  \and
Dept.\ of Electrical Engineering, Chalmers University of Technology, Gothenburg, Sweden\label{chalmers}
  \and
Department of Physics, Virginia Polytechnic Institute and State University, 50 West Campus Drive, Blacksburg, VA 24061, USA\label{virginiatech}
  \and
CSIRO Astronomy and Space Science, Australia Telescope National Facility, PO Box 76, Epping NSW 1710, Australia\label{csiro}
  \and
Sydney Institute for Astronomy, School of Physics, University of Sydney, Sydney, New South Wales 2006, Australia\label{sydney}
  \and
Rijksuniversiteit Groningen Center for Information Technology, P.O. Box 11044, 9700 CA Groningen, the Netherlands\label{rugcit}}

%% file: inc-acks.tex
This research was supported by 
the European Research Council under the European Union's Seventh Framework Programme
(FP/2007-2013)/ERC Grant Agreement No. 617199 (`ALERT'), 
and by Vici research programme `ARGO' with project number
639.043.815, financed by the Dutch Research Council (NWO). 
Instrumentation development was supported 
by NWO (grant 614.061.613 `ARTS') and the  
Netherlands Research School for Astronomy (`NOVA4-ARTS' and `NOVA-NW3').
SMS acknowledges support from the National Aeronautics and Space Administration (NASA) under grant
number NNX17AL74G issued through the NNH16ZDA001N Astrophysics Data Analysis Program (ADAP). 
DV acknowledges support
from the Netherlands eScience Center (NLeSC) under grant ASDI.15.406. 
EAKA is supported by the WISE research programme,
which is financed by NWO. 
MI acknowledges funding from the EU FP7 MCA - Swedish VINNOVA VINMER Fellowship under grant 2009-01175.

%% file: repeaters_arxiv.bbl
\begin{thebibliography}{54}
\expandafter\ifx\csname natexlab\endcsname\relax\def\natexlab#1{#1}\fi

\bibitem[{{Adams} \& {van Leeuwen}(2019)}]{2019NatAs...3..188A}
{Adams}, E. A.~K. \& {van Leeuwen}, J. 2019, Nature Astronomy, 3, 188

\bibitem[{{Aschwanden} \& {McTiernan}(2010)}]{aschwanden10}
{Aschwanden}, M.~J. \& {McTiernan}, J.~M. 2010, \apj, 717, 683

\bibitem[{{Bhat} {et~al.}(2008){Bhat}, {Tingay}, \& {Knight}}]{Bhat-2008}
{Bhat}, N.~D.~R., {Tingay}, S.~J., \& {Knight}, H.~S. 2008, \apj, 676, 1200

\bibitem[{{Chatterjee} {et~al.}(2017){Chatterjee}, {Law}, {Wharton},
  {Burke-Spolaor}, {Hessels}, {Bower}, {Cordes}, {Tendulkar}, {Bassa},
  {Demorest}, {Butler}, {Seymour}, {Scholz}, {Abruzzo}, {Bogdanov}, {Kaspi},
  {Keimpema}, {Lazio}, {Marcote}, {McLaughlin}, {Paragi}, {Ransom}, {Rupen},
  {Spitler}, \& {van Langevelde}}]{Chatterjee17}
{Chatterjee}, S., {Law}, C.~J., {Wharton}, R.~S., {et~al.} 2017, \nat, 541, 58

\bibitem[{{CHIME/FRB Collaboration} {et~al.}(2019{\natexlab{a}}){CHIME/FRB
  Collaboration}, {Amiri}, {Bandura}, {Bhardwaj}, {Boubel}, {Boyce}, {Boyle},
  {. Brar}, {Burhanpurkar}, {Cassanelli}, {Chawla}, {Cliche}, {Cubranic},
  {Deng}, {Denman}, {Dobbs}, {Fandino}, {Fonseca}, {Gaensler}, {Gilbert},
  {Gill}, {Giri}, {Good}, {Halpern}, {Hanna}, {Hill}, {Hinshaw}, {H{\"o}fer},
  {Josephy}, {Kaspi}, {Landecker}, {Lang}, {Lin}, {Masui}, {Mckinven},
  {Mena-Parra}, {Merryfield}, {Michilli}, {Milutinovic}, {Moatti}, {Naidu},
  {Newburgh}, {Ng}, {Patel}, {Pen}, {Pinsonneault-Marotte}, {Pleunis},
  {Rafiei-Ravandi}, {Rahman}, {Ransom}, {Renard}, {Scholz}, {Shaw}, {Siegel},
  {Smith}, {Stairs}, {Tendulkar}, {Tretyakov}, {Vanderlinde}, \&
  {Yadav}}]{R2CHIME19}
{CHIME/FRB Collaboration}, {Amiri}, M., {Bandura}, K., {et~al.}
  2019{\natexlab{a}}, \nat, 566, 235

\bibitem[{{CHIME/FRB Collaboration} {et~al.}(2019{\natexlab{b}}){CHIME/FRB
  Collaboration}, {Andersen}, {Band ura}, {Bhardwaj}, {Boubel}, {Boyce},
  {Boyle}, {Brar}, {Cassanelli}, {Chawla}, {Cubranic}, {Deng}, {Dobbs},
  {Fandino}, {Fonseca}, {Gaensler}, {Gilbert}, {Giri}, {Good}, {Halpern},
  {Hill}, {Hinshaw}, {H{\"o}fer}, {Josephy}, {Kaspi}, {Kothes}, {Landecker},
  {Lang}, {Li}, {Lin}, {Masui}, {Mena-Parra}, {Merryfield}, {Mckinven},
  {Michilli}, {Milutinovic}, {Naidu}, {Newburgh}, {Ng}, {Patel}, {Pen},
  {Pinsonneault-Marotte}, {Pleunis}, {Rafiei-Ravandi}, {Rahman}, {Ransom},
  {Renard}, {Scholz}, {Siegel}, {Singh}, {Smith}, {Stairs}, {Tendulkar},
  {Tretyakov}, {Vanderlinde}, {Yadav}, \& {Zwaniga}}]{CHIME19c}
{CHIME/FRB Collaboration}, {Andersen}, B.~C., {Band ura}, K., {et~al.}
  2019{\natexlab{b}}, arXiv e-prints, arXiv:1908.03507

\bibitem[{{Connor} {et~al.}(2016){Connor}, {Pen}, \&
  {Oppermann}}]{connor-2016b}
{Connor}, L., {Pen}, U.-L., \& {Oppermann}, N. 2016, \mnras, 458, L89

\bibitem[{{Connor} \& {van Leeuwen}(2018)}]{cl18}
{Connor}, L. \& {van Leeuwen}, J. 2018, \aj, 156, 256

\bibitem[{Cordes \& {McLaughlin}(2003)}]{CM03}
Cordes, J.~M. \& {McLaughlin}, M.~A. 2003, ApJ, 596, 1142

\bibitem[{{Fonseca} {et~al.}(2020){Fonseca}, {Andersen}, {Bhardwaj}, {Chawla},
  {Good}, {Josephy}, {Kaspi}, {Masui}, {Mckinven}, {Michilli}, {Pleunis},
  {Shin}, {Tendulkar}, {Bandura}, {Boyle}, {Brar}, {Cassanelli}, {Cubranic},
  {Dobbs}, {Dong}, {Gaensler}, {Hinshaw}, {Land ecker}, {Leung}, {Li}, {Lin},
  {Mena-Parra}, {Merryfield}, {Naidu}, {Ng}, {Patel}, {Pen}, {Rafiei-Ravandi},
  {Rahman}, {Ransom}, {Scholz}, {Smith}, {Stairs}, {Vanderlinde}, {Yadav}, \&
  {Zwaniga}}]{2020arXiv200103595F}
{Fonseca}, E., {Andersen}, B.~C., {Bhardwaj}, M., {et~al.} 2020, arXiv
  e-prints, arXiv:2001.03595

\bibitem[{{Gourdji} {et~al.}(2019){Gourdji}, {Michilli}, {Spitler}, {Hessels},
  {Seymour}, {Cordes}, \& {Chatterjee}}]{gourdji19}
{Gourdji}, K., {Michilli}, D., {Spitler}, L.~G., {et~al.} 2019, \apjl, 877, L19

\bibitem[{{G{\"o}{\v{g}}{\"u}{\c{s}} } {et~al.}(1999){G{\"o}{\v{g}}{\"u}{\c{s}}
  }, {Woods}, {Kouveliotou}, {van Paradijs}, {Briggs}, {Duncan}, \&
  {Thompson}}]{1999ApJ...526L..93G}
{G{\"o}{\v{g}}{\"u}{\c{s}} }, E., {Woods}, P.~M., {Kouveliotou}, C., {et~al.}
  1999, \apjl, 526, L93

\bibitem[{{Hankins} {et~al.}(2016){Hankins}, {Eilek}, \& {Jones}}]{Hankins16}
{Hankins}, T.~H., {Eilek}, J.~A., \& {Jones}, G. 2016, ApJ, 833, 47

\bibitem[{{Hessels} {et~al.}(2019){Hessels}, {Spitler}, {Seymour}, {Cordes},
  {Michilli}, {Lynch}, {Gourdji}, {Archibald}, {Bassa}, {Bower}, {Chatterjee},
  {Connor}, {Crawford}, {Deneva}, {Gajjar}, {Kaspi}, {Keimpema}, {Law},
  {Marcote}, {McLaughlin}, {Paragi}, {Petroff}, {Ransom}, {Scholz}, {Stappers},
  \& {Tendulkar}}]{hss+19}
{Hessels}, J.~W.~T., {Spitler}, L.~G., {Seymour}, A.~D., {et~al.} 2019, \apjl,
  876, L23

\bibitem[{{Johnston} \& {Romani}(2002)}]{2002MNRAS.332..109J}
{Johnston}, S. \& {Romani}, R.~W. 2002, \mnras, 332, 109

\bibitem[{{Josephy} {et~al.}(2019){Josephy}, {Chawla}, {Fonseca}, {Ng},
  {Patel}, {Pleunis}, {Scholz}, {Andersen}, {Bandura}, {Bhardwaj}, {Boyce},
  {Boyle}, {Brar}, {Cubranic}, {Dobbs}, {Gaensler}, {Gill}, {Giri}, {Good},
  {Halpern}, {Hinshaw}, {Kaspi}, {Landecker}, {Lang}, {Lin}, {Masui},
  {Mckinven}, {Mena-Parra}, {Merryfield}, {Michilli}, {Milutinovic}, {Naidu},
  {Pen}, {Rafiei-Ravand i}, {Rahman}, {Ransom}, {Renard}, {Siegel}, {Smith},
  {Stairs}, {Tendulkar}, {Vanderlinde}, {Yadav}, \& {Zwaniga}}]{CHIMER1}
{Josephy}, A., {Chawla}, P., {Fonseca}, E., {et~al.} 2019, \apjl, 882, L18

\bibitem[{{Karuppusamy} {et~al.}(2010){Karuppusamy}, {Stappers}, \& {van
  Straten}}]{Karuppusamy-2010}
{Karuppusamy}, R., {Stappers}, B.~W., \& {van Straten}, W. 2010, \aap, 515, A36

\bibitem[{{Law} {et~al.}(2017){Law}, {Abruzzo}, {Bassa}, {Bower},
  {Burke-Spolaor}, {Butler}, {Cantwell}, {Carey}, {Chatterjee}, {Cordes},
  {Demorest}, {Dowell}, {Fender}, {Gourdji}, {Grainge}, {Hessels}, {Hickish},
  {Kaspi}, {Lazio}, {McLaughlin}, {Michilli}, {Mooley}, {Perrott}, {Ransom},
  {Razavi-Ghods}, {Rupen}, {Scaife}, {Scott}, {Scholz}, {Seymour}, {Spitler},
  {Stovall}, {Tendulkar}, {Titterington}, {Wharton}, \& {Williams}}]{law17}
{Law}, C.~J., {Abruzzo}, M.~W., {Bassa}, C.~G., {et~al.} 2017, \apj, 850, 76

\bibitem[{{Lin} \& {Sang}(2019)}]{lin19}
{Lin}, H.-N. \& {Sang}, Y. 2019, \mnras, 2746

\bibitem[{{Lorimer} {et~al.}(2007){Lorimer}, {Bailes}, {McLaughlin},
  {Narkevic}, \& {Crawford}}]{Lorimer07}
{Lorimer}, D.~R., {Bailes}, M., {McLaughlin}, M.~A., {Narkevic}, D.~J., \&
  {Crawford}, F. 2007, Science, 318, 777

\bibitem[{{Lundgren} {et~al.}(1995){Lundgren}, {Cordes}, {Ulmer}, {Matz},
  {Lomatch}, {Foster}, \& {Hankins}}]{L95}
{Lundgren}, S.~C., {Cordes}, J.~M., {Ulmer}, M., {et~al.} 1995, ApJ, 453, 433

\bibitem[{{Maan}(2015)}]{Maan15}
{Maan}, Y. 2015, ApJ, 815, 126

\bibitem[{{Maan} \& {Aswathappa}(2014)}]{MA14}
{Maan}, Y. \& {Aswathappa}, H.~A. 2014, MNRAS, 445, 3221

\bibitem[{{Maan} {et~al.}(2019){Maan}, {Joshi}, {Surnis}, {Bagchi}, \&
  {Manoharan}}]{Maan19b}
{Maan}, Y., {Joshi}, B.~C., {Surnis}, M.~P., {Bagchi}, M., \& {Manoharan},
  P.~K. 2019, ApJL, 882, L9

\bibitem[{{Maan} \& {van Leeuwen}(2017)}]{2017arXiv170906104M}
{Maan}, Y. \& {van Leeuwen}, J. 2017, IEEE Proc. URSI GASS
  [\eprint[arXiv]{1709.06104}]

\bibitem[{{Marcote} {et~al.}(2017){Marcote}, {Paragi}, {Hessels}, {Keimpema},
  {van Langevelde}, {Huang}, {Bassa}, {Bogdanov}, {Bower}, {Burke-Spolaor},
  {Butler}, {Campbell}, {Chatterjee}, {Cordes}, {Demorest}, {Garrett}, {Ghosh},
  {Kaspi}, {Law}, {Lazio}, {McLaughlin}, {Ransom}, {Salter}, {Scholz},
  {Seymour}, {Siemion}, {Spitler}, {Tendulkar}, \& {Wharton}}]{Marcote17}
{Marcote}, B., {Paragi}, Z., {Hessels}, J.~W.~T., {et~al.} 2017, \apjl, 834, L8

\bibitem[{{Metzger} {et~al.}(2019){Metzger}, {Margalit}, \&
  {Sironi}}]{metzger19}
{Metzger}, B.~D., {Margalit}, B., \& {Sironi}, L. 2019, \mnras, 485, 4091

\bibitem[{{Michilli} {et~al.}(2018){Michilli}, {Seymour}, {Hessels}, {Spitler},
  {Gajjar}, {Archibald}, {Bower}, {Chatterjee}, {Cordes}, {Gourdji}, {Heald},
  {Kaspi}, {Law}, {Sobey}, {Adams}, {Bassa}, {Bogdanov}, {Brinkman},
  {Demorest}, {Fernand ez}, {Hellbourg}, {Lazio}, {Lynch}, {Maddox}, {Marcote},
  {McLaughlin}, {Paragi}, {Ransom}, {Scholz}, {Siemion}, {Tendulkar}, {van
  Rooy}, {Wharton}, \& {Whitlow}}]{msh+18}
{Michilli}, D., {Seymour}, A., {Hessels}, J.~W.~T., {et~al.} 2018, \nat, 553,
  182

\bibitem[{{Mickaliger} {et~al.}(2012){Mickaliger}, {McLaughlin}, {Lorimer},
  {Langston}, {Bilous}, {Kondratiev}, {Lyutikov}, {Ransom}, \&
  {Palliyaguru}}]{2012ApJ...760...64M}
{Mickaliger}, M.~B., {McLaughlin}, M.~A., {Lorimer}, D.~R., {et~al.} 2012,
  \apj, 760, 64

\bibitem[{{Mikhailov}(2018)}]{mikh18}
{Mikhailov}, K. 2018, PhD thesis, University of Amsterdam, {Ch.~4,
  \url{http://hdl.handle.net/11245.1/d3a5406f-eb36-4bb8-a280-54a3eedd0a52}}

\bibitem[{{Oosterloo} {et~al.}(2010){Oosterloo}, {Verheijen}, \& {van
  Cappellen}}]{ovc10}
{Oosterloo}, T., {Verheijen}, M., \& {van Cappellen}, W. 2010, in "ISKAF2010
  Science Meeting", Van Leeuwen, Morganti, Serra (Eds.)

\bibitem[{{Oostrum} {et~al.}(2017){Oostrum}, {van Leeuwen}, {Attema}, {van
  Cappellen}, {Connor}, {Hut}, {Maan}, {Oosterloo}, {Petroff}, {van der
  Schuur}, {Sclocco}, \& {Verheijen}}]{atel121102}
{Oostrum}, L.~C., {van Leeuwen}, J., {Attema}, J., {et~al.} 2017, The
  Astronomer's Telegram, 10693, 1

\bibitem[{{Oostrum} {et~al.}(2020){Oostrum}, {van Leeuwen}, {Maan}, {Coenen},
  \& {Ishwara-Chandra}}]{2020MNRAS.tmp..131O}
{Oostrum}, L.~C., {van Leeuwen}, J., {Maan}, Y., {Coenen}, T., \&
  {Ishwara-Chandra}, C.~H. 2020, \mnras, 131

\bibitem[{{Oppermann} {et~al.}(2018){Oppermann}, {Yu}, \& {Pen}}]{orp18}
{Oppermann}, N., {Yu}, H.-R., \& {Pen}, U.-L. 2018, \mnras, 475, 5109

\bibitem[{{Patek} \& {CHIME/FRB Collaboration}(2019)}]{Patek19}
{Patek}, C. \& {CHIME/FRB Collaboration}. 2019, The Astronomer's Telegram,
  13013, 1

\bibitem[{{Pearlman} {et~al.}(2018){Pearlman}, {Majid}, {Prince}, {Kocz}, \&
  {Horiuchi}}]{Pearlman18}
{Pearlman}, A.~B., {Majid}, W.~A., {Prince}, T.~A., {Kocz}, J., \& {Horiuchi},
  S. 2018, ApJ, 866, 160

\bibitem[{{Perley} \& {Butler}(2017)}]{pb17}
{Perley}, R.~A. \& {Butler}, B.~J. 2017, \apjs, 230, 7

\bibitem[{{Petroff} {et~al.}(2019){Petroff}, {Hessels}, \& {Lorimer}}]{phl19}
{Petroff}, E., {Hessels}, J.~W.~T., \& {Lorimer}, D.~R. 2019, \aapr, 27, 4

\bibitem[{{Petroff} {et~al.}(2015){Petroff}, {Johnston}, {Keane}, {van
  Straten}, {Bailes}, {Barr}, {Barsdell}, {Burke-Spolaor}, {Caleb}, {Champion},
  {Flynn}, {Jameson}, {Kramer}, {Ng}, {Possenti}, \& {Stappers}}]{Petroff2015}
{Petroff}, E., {Johnston}, S., {Keane}, E.~F., {et~al.} 2015, \mnras, 454, 457

\bibitem[{{Piro} \& {Gaensler}(2018)}]{piro18}
{Piro}, A.~L. \& {Gaensler}, B.~M. 2018, \apj, 861, 150

\bibitem[{{Ramkumar} \& {Deshpande}(1999)}]{ramkumar99}
{Ramkumar}, P.~S. \& {Deshpande}, A.~A. 1999, Journal of Astrophysics and
  Astronomy, 20, 37

\bibitem[{{Ransom}(2011)}]{presto}
{Ransom}, S. 2011, {PRESTO: PulsaR Exploration and Search TOolkit}

\bibitem[{{Scholz} {et~al.}(2016){Scholz}, {Spitler}, {Hessels}, {Chatterjee},
  {Cordes}, {Kaspi}, {Wharton}, {Bassa}, {Bogdanov}, {Camilo}, {Crawford},
  {Deneva}, {van Leeuwen}, {Lynch}, {Madsen}, {McLaughlin}, {Mickaliger},
  {Parent}, {Patel}, {Ransom}, {Seymour}, {Stairs}, {Stappers}, \&
  {Tendulkar}}]{scholz16}
{Scholz}, P., {Spitler}, L.~G., {Hessels}, J.~W.~T., {et~al.} 2016, \apj, 833,
  177

\bibitem[{{Sclocco} {et~al.}(2016){Sclocco}, {van Leeuwen}, {Bal}, \& {van
  Nieuwpoort}}]{slb+16}
{Sclocco}, A., {van Leeuwen}, J., {Bal}, H.~E., \& {van Nieuwpoort}, R.~V.
  2016, Astronomy and Computing, 14, 1

\bibitem[{{Serylak} {et~al.}(2009){Serylak}, {Stappers}, {Weltevrede},
  {Kramer}, {Jessner}, {Lyne}, {Jordan}, {Lazaridis}, \& {Zensus}}]{Serylak09}
{Serylak}, M., {Stappers}, B.~W., {Weltevrede}, P., {et~al.} 2009, MNRAS, 394,
  295

\bibitem[{{Spitler} {et~al.}(2016){Spitler}, {Scholz}, {Hessels}, {Bogdanov},
  {Brazier}, {Camilo}, {Chatterjee}, {Cordes}, {Crawford}, {Deneva}, {Ferdman},
  {Freire}, {Kaspi}, {Lazarus}, {Lynch}, {Madsen}, {McLaughlin}, {Patel},
  {Ransom}, {Seymour}, {Stairs}, {Stappers}, {van Leeuwen}, \&
  {Zhu}}]{spitler16}
{Spitler}, L.~G., {Scholz}, P., {Hessels}, J.~W.~T., {et~al.} 2016, \nat, 531,
  202

\bibitem[{{Straal}(2018)}]{2018PhDT.......155S}
{Straal}, S.~M. 2018, PhD thesis, Anton Pannekoek Institute for Astronomy,
  University of Amsterdam, Science Park 904, 1098 XH, Amsterdam, The
  Netherlands

\bibitem[{{Tendulkar} {et~al.}(2017){Tendulkar}, {Bassa}, {Cordes}, {Bower},
  {Law}, {Chatterjee}, {Adams}, {Bogdanov}, {Burke-Spolaor}, {Butler},
  {Demorest}, {Hessels}, {Kaspi}, {Lazio}, {Maddox}, {Marcote}, {McLaughlin},
  {Paragi}, {Ransom}, {Scholz}, {Seymour}, {Spitler}, {van Langevelde}, \&
  {Wharton}}]{Tendulkar-2017}
{Tendulkar}, S.~P., {Bassa}, C.~G., {Cordes}, J.~M., {et~al.} 2017, \apjl, 834,
  L7

\bibitem[{{Thornton} {et~al.}(2013){Thornton}, {Stappers}, {Bailes},
  {Barsdell}, {Bates}, {Bhat}, {Burgay}, {Burke-Spolaor}, {Champion}, {Coster},
  {D'Amico}, {Jameson}, {Johnston}, {Keith}, {Kramer}, {Levin}, {Milia}, {Ng},
  {Possenti}, \& {van Straten}}]{Thornton13}
{Thornton}, D., {Stappers}, B., {Bailes}, M., {et~al.} 2013, Science, 341, 53

\bibitem[{{van Leeuwen}(2014)}]{leeu14}
{van Leeuwen}, J. 2014, in The Third Hot-wiring the Transient Universe
  Workshop, ed. P.~R. {Wozniak}, M.~J. {Graham}, A.~A. {Mahabal}, \&
  R.~{Seaman}, 79--79

\bibitem[{{van Leeuwen} {et~al.}(2019){van Leeuwen}, {Mikhailov}, {Keane},
  {Coenen}, {Connor}, {Kondratiev}, {Michilli}, \& {Sanidas}}]{lkk+19}
{van Leeuwen}, J., {Mikhailov}, K., {Keane}, E., {et~al.} 2019, \aap,
  arXiv:1911.11228, {\it accepted}

\bibitem[{{Wadiasingh} {et~al.}(2019){Wadiasingh}, {Beniamini}, {Timokhin},
  {Baring}, {van der Horst}, {Harding}, \& {Kazanas}}]{2019arXiv191006979W}
{Wadiasingh}, Z., {Beniamini}, P., {Timokhin}, A., {et~al.} 2019, arXiv
  e-prints, arXiv:1910.06979

\bibitem[{{Wadiasingh} \& {Timokhin}(2019)}]{2019ApJ...879....4W}
{Wadiasingh}, Z. \& {Timokhin}, A. 2019, \apj, 879, 4

\bibitem[{{Wheatland}(2000)}]{wheatland00}
{Wheatland}, M.~S. 2000, \apjl, 536, L109

\end{thebibliography}
